
\def\underarrow#1{\vbox{\ialign{##\crcr$\hfil\displaystyle
{#1}\hfil$\crcr\noalign{\kern1pt
\nointerlineskip}$\longrightarrow$\crcr}}}
%
\def\d{{\rm d}}
\def\ltorder{\mathrel{\raise.3ex\hbox{$<$}\mkern-14mu
             \lower0.6ex\hbox{$\sim$}}}
\def\lesssim{\mathrel{\raise.3ex\hbox{$<$}\mkern-14mu
             \lower0.6ex\hbox{$\sim$}}}
\def\Z{{\bf Z}}
\def\C{{\bf C}}
\def\R{{\bf R}}

\def\p{\partial}
\def\to{\rightarrow}
\def\d{{\rm d}}
\def\catquot{\smash{\mathrel{/\!/}}}
\input phyzzx
\overfullrule=0pt
\tolerance=5000
\twelvepoint

\pubnum{IASSNS-HEP-93/3}
\date{January, 1993}
\titlepage
\title{PHASES OF $N=2$ THEORIES IN TWO DIMENSIONS}
\vglue-.25in
\author{Edward Witten
\foot{Research supported in part by NSF Grant
PHY92-45317.}}
\medskip
\address{School of Natural Sciences
\break Institute for Advanced Study
\break Olden Lane
\break Princeton, NJ 08540}
\bigskip
\abstract{By looking at phase transitions which occur as parameters
are varied in supersymmetric gauge theories, a natural relation is found
between sigma models based on Calabi-Yau hypersurfaces in weighted projective
spaces and Landau-Ginzburg models.  The construction permits one to recover
the known correspondence between these types of models
and to greatly extend it to include new classes of
manifolds and also to include models with $(0,2)$ world-sheet supersymmetry.
The construction also predicts the possibility of certain
physical processes involving a change in the topology of space-time.}

\def\b{\overline}
\def\CP{{\bf CP}}
\def\C{{\bf C}}

\endpage
\chapter{INTRODUCTION}

The present paper is based on systematically exploiting one simple idea
which is familiar in $N=1$
supersymmetric theories in four dimensions and which we will therefore
first state in that context.
We consider renormalizable gauge theories constructed from vector (gauge)
multiplets and charged chiral multiplets.
If the gauge group includes a $U(1)$ factor, one of the possible
supersymmetric interactions is the Fayet-Iliopoulos $D$ term.  In
terms of a vector superfield $V$, this term can be written as
$$ -r\int \d^4x\,\,\d^4\theta \,\,\,\,\,\,V.  \eqn\oppo$$
The potential energy for scalar components
$s_i, \,\,i=1\dots k$, of chiral multiplets $S_i$ of charge $n_i$
is then
$$U(s_i) ={1\over 2e^2}D^2 +\sum_i\left|{\partial W\over\partial s_i}\right|^2,
\eqn\omigo$$
where $e$ is the gauge coupling, $W$ is a holomorphic function known
as the superpotential, and
$$ D=-e^2(\sum_in_i|s_i|^2-r).   \eqn\oopigo$$
Actually, $D$ can be interpreted as the Hamiltonian function for the $U(1)$
action on  a copy of $\C^k$ (on which the $s_i$ are coordinates and which
we endow with the Kahler form $\omega=-i\sum_j\d s_j\wedge\d\bar s_j$).
The parameter $r$ corresponds to the familiar possibility of adding
a constant to the Hamiltonian.

As $r$ is varied, such a system will typically undergo phase transitions,
in many cases leaving supersymmetry unbroken but otherwise changing
the pattern of massless fields.
This is a familiar and important story in phenomenology of renormalizable
supersymmetric theories in four dimensions.
Our intention in the present paper is to study the same models,
but dimensionally reduced to $D=2$.  The two dimensional
version of the story is particularly rich, because milder anomaly
cancellation conditions give more freedom in constructing models
and because of the usual
special properties of massless fields in two dimensions.

\REF\martinec{E.Martinec, ``Criticality, Catastrophes, and
Compactifications,'' in {\it Physics And Mathematics Of Strings},
ed L. Brink, D. Friedan, and A. M. Polyakov (World Scientific, 1990).}
\REF\vafa{C. Vafa and N. Warner, ``Catastrophes And The Classification
Of Conformal Field Theories,'' Phys. Lett. {\bf 218B} (1989) 51.}
\REF\greene{B. Greene, C. Vafa, and N. Warner, ``Calabi-Yau Manifolds
And Renormalization Group Flows,'' Nucl. Phys. {\bf B324} (1989) 371.}
\REF\lvw{W. Lerche, C. Vafa, and N. Warner, ``Chiral Rings In $N=2$
Superconformal Theory,'' Nucl. Phys. {\bf B324} (1989) 427.}
\REF\cecgp{S. Cecotti, L. Girardello, and A. Pasquinucci,
``Nonperturbative Aspects And Exact Results In The $N=2$
Landau-Ginzburg Models,'' Nucl. Phys. {\bf B328} (1989) 701,
``Singularity Theory And $N=2$ Supersymmetry,'' Int. J. Mod. Phys.
{\bf A6} (1991) 2427.}
\REF\cecotti{S. Cecotti, ``N=2 Landau-Ginzburg vs. Calabi-Yau $\sigma$-Models:
Non-Perturbative Aspects,'' Int. J. Mod. Phys. {\bf A6} (1991) 1749.}
As a special case which originally motivated this investigation,
we will recover the familiar correspondence between
Calabi-Yau and Landau-Ginzburg models [\martinec,\vafa,\greene].
The usual arguments for this correspondence are heuristic arguments
of universality [\martinec--\greene] including the simple fact
that the Calabi-Yau and Landau-Ginzburg models are $N=2$ models
parametrized by the same data,
a heuristic path integral argument [\greene], and
equivalence of the $cc$ chiral rings, which is
discussed in the papers just cited
and in [\lvw -\cecotti].
{}From our construction we will recapture essentially all the concrete
evidence for the Calabi-Yau/Landau-Ginzburg correspondence along
with new features.
The argument will be carried out without assuming
or proving conformal invariance of either Calabi-Yau or Landau-Ginzburg
models; the picture is certainly richer when supplemented by
(well-known) independent arguments for conformal invariance.

We will also find many generalizations of the usual C-Y/L-G correspondence,
involving more general classes of Calabi-Yau manifolds
(hypersurfaces in products of projective spaces and in more general
toric and other varieties, and intersections of such hypersurfaces).
One novelty is that in general one must consider gauged Landau-Ginzburg models
as well as ordinary ones.  Moreover, we will get
an extension of the C-Y/L-G correspondence to
models with $(0,2)$ world-sheet
supersymmetry.  Hitherto only the case of $(2,2)$
supersymmetry has been considered.  $(0,2)$ world-sheet supersymmetry
is of phenomenological interest as it can naturally lead, for instance,
to models with space-time grand unified gauge group $SU(5)$
or $SO(10)$ rather than $E_6$.

\REF\amg{P. Aspinwall, B. Greene, and D. Morrison, ``Multiple
Mirror Manifolds And Topology Change In String Theory,'' IAS preprint
(1993).}
Variation of the coefficients of the Fayet-Iliopoulos $D$ terms can
also induce phase transitions involving
changes in the topology of a Calabi-Yau manifold.  This will be investigated
from a local point of view only in \S5.5.  For reasons explained in \S4,
the topology changes in question preserve the birational equivalence
class of the Calabi-Yau manifold.  Precisely this situation
has been studied globally by Aspinwall, Greene, and Morrison
[\amg] who by using mirror symmetry (and finding pairs of
topologically different but birationally
equivalent manifolds with the same mirror) obtained
very precise and detailed evidence for the possibility of physical
processes with change of topology.  The present problem turned out
to have an unexpectedly close relation  to their investigation,
and I benefited from discussions with them, especially in \S5.5.
The results of [\amg] and of \S5.5 are the first concrete evidence
for the (long-suspected) occurrence in string theory of processes
with change of the space-time topology.

\REF\hubscho{T. Hubsch, ``Chameleonic $\sigma$-Models,'' Phys.
Lett. {\bf 247B} (1990) 317, ``How Singular A Space Can Superstrings
Thread?'' Mod. Phys. Lett. {\bf A6} (1991) 207; J. Gates and T. Hubsch,
``Unidexterous Locally Supersymmetric Actions For Calabi-Yau
Compactifications,'' Phys. Lett. {\bf 226B} (1989) 100,
``Calabi-Yau Heterotic Strings And Supersymmetric Sigma Models,''
Nucl. Phys. {\bf B343} (1990) 741.}
The present paper is based on familiar representations
of the relevant sigma models [\hubscho], except that we will consider
linear sigma models (rather than the nonlinear ones to which they reduce
at low energies);  this gives a useful added freedom of maneuver.

\REF\thaddeus{M. Thaddeus, ``Stable Pairs, Linear Systems, And
The Verlinde Formula,'' MSRI preprint (1992).}
While this work was in progress, I received a paper by Thaddeus
[\thaddeus] who uses a mathematical idea closely related to that of the
present paper
to prove the Verlinde formula for the dimension of the space
of conformal blocks of the $SU(2)$ WZW model.

\chapter{FIELD THEORY BACKGROUND}

In this section, I will present the field theory background
to our subsequent analysis.
Most of the detailed formulas of this section are not needed for
reading the paper, and are presented for reference
and to be more self-contained.

\REF\wess{J. Wess and J. Bagger, {\it Supersymmetry And Supergravity},
Princeton University Press (second edition, 1992).}
\REF\rsbook{S. J. Gates, Jr., M. T. Grisaru, M. Rocek, and W. Siegel,
{\it Superspace, or One Thousand And One Lessons In Supersymmetry},
(Benjamin-Cummings, 1983).}
\REF\freund{P. G. O. Freund, {\it Introduction To Supersymmetry}
(Camb. Univ. Press, 1986).}
\REF\west{P. West, {\it Introduction To Supersymmetry And Supergravity}
(World Scientific, 1986).}
\REF\misra{S. P. Misra, {\it Introduction To Supersymmetry And
Supergravity} (Wiley Eastern Limited, 1992).}
$N=2$ supersymmetry in two dimensions can be obtained by dimensional
reduction from $N=1$ supersymmetry in four dimensional space-time.
A number  of good books exist [\wess--\misra].  Our conventions for four
dimensional superfields will be those of [\wess].

In superspace with coordinates $x^m,\theta^\alpha,\b\theta^{\dot\alpha}$,
supersymmetry is realized geometrically by the operators
$$\eqalign{Q_\alpha & = {\partial\over\partial\theta^\alpha}-i\sigma^m_{\alpha
\dot\alpha}\b\theta^{\dot\alpha}{\partial\over\partial x^m} \cr
\b Q_{\dot\alpha} & =- {\partial\over\partial\b\theta^{\dot\alpha}}
+i\sigma^m_{\alpha
\dot\alpha}\theta^{\alpha}{\partial\over\partial x^m} \cr}\eqn\suge$$
$\alpha$ and $\dot\alpha$ are the two chiralities of spinor indices;
also one writes $\psi_\alpha=\epsilon_{\alpha\beta}\psi^\beta,\;
\psi^{\alpha}=\epsilon^{\alpha\beta}\psi_{\beta}$,
where $\epsilon$ is the antisymmetric tensor with $\epsilon^{12}
=-\epsilon_{1 2}=1$; and similarly for dotted indices.
The tensors $\sigma^m_{\alpha\dot\alpha}$ are in the representation
used in [\wess]
$$\eqalign{\sigma^0 & = \left(\matrix{ -1 & 0 \cr 0 & -1 \cr}\right) \cr
\sigma^1 & = \left(\matrix{ 0  & 1 \cr 1 &  0 \cr}\right) \cr
\sigma^2 & = \left(\matrix{ 0 & -i  \cr i & 0  \cr}\right) \cr
\sigma^3 & = \left(\matrix{ 1 & 0 \cr 0 & -1 \cr}\right) \cr}\eqn\nuvo$$
The metric is $\eta^{mn}={\rm diag}(-1,1,1,1)$.
The supersymmetry generators of equation \suge\ commute with the
operators
$$\eqalign{D_\alpha & = {\partial\over\partial\theta^\alpha}+i\sigma^m_{\alpha
\dot\alpha}\b\theta^{\dot\alpha}{\partial\over\partial x^m} \cr
\b D_{\dot\alpha} & =- {\partial\over\partial\b\theta^{\dot\alpha}}
-i\sigma^m_{\alpha
\dot\alpha}\theta^{\alpha}{\partial\over\partial x^m} \cr}\eqn\suge$$
which are used in writing Lagrangians.

The simplest type of superfield is a chiral superfield $\Phi$ which obeys
$\b D_{\dot\alpha}\Phi=0$ and can be expanded
$$\Phi(x,\theta)=\phi(y)+\sqrt 2 \theta^\alpha\psi_\alpha(y)
+\theta^\alpha\theta_\alpha F(y),
\eqn\momo$$
where $y^m=x^m+i\theta^\alpha
\sigma^m_{\alpha\dot\alpha}\b\theta^{\dot\alpha}$.
The complex conjugate of $\Phi$ is an antichiral multiplet $\b\Phi$,
obeying $D_\alpha\b\Phi=0$, and with an expansion
$$\b\Phi  =\b \phi(\b y)+\sqrt 2\b\theta_{\dot\alpha}\b\psi^{\dot\alpha}(\b y)
+\b\theta_{\dot\alpha}\b\theta^{\dot\alpha}F(\b y), \eqn\copo$$
where ${\b y}^m=x^m-i\theta^\alpha\sigma^m_{\alpha\dot\alpha}\b\theta^{\dot
\alpha}$.  In general, the $\theta=0$ component of a chiral field
labeled by a capital letter $\Phi,S,P,\dots$ will be denoted by the
corresponding lower case letter $\phi,s,p,\dots$.

\def\cd{{\cal D}}
To formulate gauge theory, one introduces
a gauge field in superspace, replacing the differential operators
$D_\alpha$, $\b D_{\dot \alpha}$, and $\partial_m=\partial/\partial x^m$
by gauge covariant derivatives $\cd_\alpha$, $\b\cd_{\dot\alpha}$,
and $\cd_m$.  One imposes however a severe restriction on the superspace
gauge fields.  To permit charged chiral superfields to exist,
one needs the integrability of the equation $\b \cd_{\dot\alpha}\Phi=0$.
So one requires
$$0=\{\b\cd_{\dot\alpha},\b\cd_{\dot\beta}\}=\{\cd_\alpha,\cd_\beta\}.
        \eqn\huju$$
These conditions ensure that, with a suitable partial gauge fixing,
$$\eqalign{\cd_\alpha & = e^{-V}\cd_\alpha e^{V} \cr
           \b\cd_{\dot\alpha} &= e^{V}\b \cd_{\dot\alpha}e^{-V},\cr}
          \eqn\hococ$$
where $V$ is a real Lie algebra valued function on superspace;
$V$ is known as a vector superfield.
One also imposes the further constraint
$$\{\cd_\alpha,\b\cd_{\dot\alpha}\}=-2i\sigma^m_{\alpha\dot\alpha}\cd_m,
\eqn\urgo$$
so that the superspace gauge field is entirely determined in terms
of $V$.

There is still a residual gauge invariance, which for gauge
group $U(1)$ -- in which case $V$ is simply a single real function on
superspace -- takes the form
$$V\to V+i(\Lambda-\b\Lambda), \eqn\pomo$$
$\Lambda$ being a chiral superfield.
A chiral superfield $\Phi$
of charge $Q$ transforms under this residual gauge invariance as
$$\Phi\to \exp(-iQ\Lambda)\cdot \Phi. \eqn\bomo$$
One can partially fix this residual  gauge invariance
(by going to what is called
Wess-Zumino gauge) to put $V$ in the form
$$V = -\theta^\alpha \sigma^m_{\alpha\dot\alpha}\b\theta^{\dot\alpha}v_m
+i\theta^\alpha\theta_\alpha \b\theta_{\dot\alpha}\b\lambda^{\dot\alpha}
-i\b\theta_{\dot\alpha}\b\theta^{\dot\alpha}\theta^\alpha\lambda_\alpha
+{1\over 2}\theta^\alpha\theta_\alpha\b\theta_{\dot\alpha}\b\theta^{\dot\alpha}
D. \eqn\urmo$$
In Wess-Zumino gauge, one still retains the ordinary gauge invariance.
This corresponds to $\Lambda=-a(x)$ (a real function of $x$ only) with
the usual transformation laws $\Phi\to \exp(iQa)\Phi$, $v\to v-\d a$.

In four dimensions, the basic
gauge invariant field strength of the abelian gauge field is
$[\cd_\alpha,\cd_m]$.
In two dimensions there is a more basic invariant that we will introduce
later.

In Wess-Zumino gauge, though the physical content of the theory is
relatively transparent, the supersymmetry transformation laws are relatively
complicated because supersymmetry transformations (generated by $Q_\alpha$
and $\b Q_{\dot\alpha}$) must be accompanied by gauge transformations to
preserve the Wess-Zumino gauge.  The transformation laws become still
more complicated when dimensionally reduced to two dimensions.
\foot{All conventions will be those that follow by dimensional reduction
from four dimensional conventions of [\wess].  I adopted this approach
because the models we will study arise so naturally by reduction from four
dimensions.  Also, there do not seem to be any standard conventions for
component expansions of
vector superfields in two dimensions.  Chiral superfields in two dimensions
are often described by conventions that differ from the present ones
by factors of $\sqrt 2 $; these factors
can be absorbed in rescaling the fields in a fairly obvious way.}
I will now write down the reduced transformation laws, in
Wess-Zumino gauge, for reference,
but I hasten to reassure the reader that this paper
can be read without detailed familiarity with the following formulas.

In making the reduction,
I will take the fields to be independent
of $x^1$ and $x^2$.
Thus, the components $v_1$ and $v_2$ of the gauge field in the $x^1$ and $x^2$
directions, along with all the other fields, are functions of $x^0,x^3$ only;
I write $\sigma=(v_1-iv_2)/\sqrt 2$, $\b\sigma=(v_1+iv_2)/\sqrt 2$.
For the remaining space-time coordinates, I will adopt a two dimensional
notation, writing $x^0=y^0, \; x^3=y^1$.
After dimensional reduction, it is convenient to label the fermion
components as
$(\psi^1,\psi^2)=(\psi^-,\psi^+)$ and $(\psi_1,\psi_2)=(\psi_-,\psi_+)$
(so $\psi^-=\psi_+$, $\psi^+=-\psi_-$), and similarly for dotted
components.
The dimensionally reduced transformation laws of the vector multiplet
under a supersymmetry transformation with parameters $\epsilon_{\pm},
\b\epsilon_{\pm}$
are (by reducing the formulas on p. 50 of [\wess])
$$\eqalign{\delta v_m & = i\b\epsilon\sigma_m \lambda +i\epsilon \sigma_m
      \b\lambda \cr
\delta\sigma & = -i\sqrt 2
         \b\epsilon_+\lambda_--i\sqrt 2\epsilon_-\b\lambda_+\cr
           \delta \b\sigma & = -i\sqrt 2\epsilon_+\b\lambda_--i\sqrt 2
                 \b\epsilon_-\lambda_+ \cr
          \delta D & =-\b\epsilon_+(\partial_0-\partial_1)\lambda_+
       -\b\epsilon_-(\partial_0+\partial_1)\lambda_-
+ \epsilon_+(\partial_0-\partial_1)\b\lambda_++\epsilon_-(\partial_0+
\partial_1)\b\lambda_- \cr
\delta\lambda_+ & = i\epsilon_+D +\sqrt 2(\partial_0+\partial_1)
\b\sigma\epsilon_-
       -v_{01}\epsilon_+ \cr
\delta\lambda_- & = i\epsilon_-D+\sqrt 2(\partial_0-\partial_1)\sigma\epsilon_+
       +v_{01}\epsilon_- \cr
\delta\b\lambda_+ & = -i\b\epsilon_+D +\sqrt 2(\partial_0+\partial_1)\sigma\b
\epsilon_-
       -v_{01}\b\epsilon_+ \cr
\delta\b\lambda_- & = -i\b\epsilon_-D
+\sqrt 2(\partial_0-\partial_1)\b\sigma\,\b\epsilon_+
       +v_{01}\b\epsilon_-, \cr
       }\eqn\donkey$$
with $v_{01}=\partial_0v_1-\partial_1v_0$.

In the presence of gauge fields, a chiral superfield $\Phi$ in a given
representation of the gauge group is a
superfield obeying  $\b\cd_{\dot\alpha}\Phi=0$, where the covariant
derivative in $\b\cd_{\dot\alpha}$ is taken in the appropriate
representation.  If we write
$$\Phi=e^V\Phi_0, \eqn\convwrite$$
then $\Phi_0$ obeys $\b D_{\dot \alpha}\Phi_0=0$ and has
a theta expansion of the form given in \momo.  The transformation laws
for these component fields come out to be
$$\eqalign{
  \delta \phi & = \sqrt 2\left(\epsilon_+\psi_--\epsilon_-\psi_+\right)\cr
  \delta \psi_+ & = i\sqrt 2(D_0+D_1)\phi\b\epsilon_-+\sqrt 2 \epsilon_+F
             -2 Q\phi \b\sigma\,\b\epsilon_+ \cr
  \delta \psi_- & =  -i\sqrt 2(D_0-D_1)\phi\b\epsilon_++\sqrt 2 \epsilon_-F
             +2 Q\phi \sigma\,\b\epsilon_- \cr
  \delta F & = -i\sqrt 2 \b\epsilon_+(D_0-D_1)\psi_+
               -i\sqrt 2 \b\epsilon_-(D_0+D_1)\psi_-
\cr &\;\;\;\;\;
 +2Q(\b\epsilon_+ \b\sigma\psi_-+\b\epsilon_-\sigma\psi_+)
          +2iQ\phi(\b\epsilon_-\b\lambda_+-\b\epsilon_+\b\lambda_-).\cr}
\eqn\ppss$$
The transformation laws of the antichiral multiplet $\b\Phi$ are the
complex conjugate of these.

\subsection{Twisted Chiral Superfields; Gauge Field Strength}

\REF\ggates{J. Gates, ``Superspace Formulation Of New Non-Linear
Sigma Models,'' Nucl. Phys. {\bf B238} (1984) 349.}
\REF\rocek{S. J. Gates, C. M. Hull, and M. Rocek, ``Twisted Multiplets
And New Supersymmetric Nonlinear $\sigma$-Models,'' Nucl. Phys.
{\bf B248} (1984) 157}
\REF\busch{
T. Buscher, U. Lindstrom, and M. Rocek, ``New Supersymmetric $\sigma$-Models
With Wess-Zumino Term,' Phys. Lett. {\bf 202B} (1988) 94.}
\REF\hitch{N. J. Hitchin, A. Karlhede, U. Lindstrom, and
M. Rocek, ``Hyper-Kahler Metrics And Supersymmetry,''
Commun. Math. Phys. {\bf 108} (1987) 535.}
\REF\newrocek{M. Rocek, ``Modified Calabi-Yau Manifolds With Torsion,''
in S.-T. Yau, ed., {\it Essays On Mirror Manifolds} (International Press,
1992).}
One of the novelties that appears in two dimensions, relative to four,
is that in addition to chiral superfields, obeying $\b D_+\Phi
=\b D_-\Phi=0$, it is possible to have twisted chiral superfields,
obeying $\b D_+\Sigma =D_-\Sigma = 0$ [\ggates--\newrocek].
Sigma models containing both chiral and twisted chiral superfields
are quite lovely.  Since mirror symmetry
turns chiral multiplets into twisted chiral multiplets,
it is likely that consideration of appropriate models containing multiplets
of both types is helpful for understanding mirror symmetry.

\REF\rocverl{M. Rocek and E. Verlinde, ``Duality, Quotients, and
Currents,'' Nucl. Phys. {\bf B373} (1992) 630.}
More important for our present purposes, however, is the fact
(exploited in [\rocverl] in explaining $N=2$ duality as an abelian
mirror symmetry) that in two dimensions the basic gauge invariant
field strength of the superspace gauge field is a twisted chiral superfield.
This quantity is
$$  \Sigma ={1\over 2\sqrt 2}\{\b\cd_+,\cd_-\}. \eqn\huco$$
which according to the Bianchi identities is annihilated by $\b\cd_+$
and $\cd_-$.
In the abelian case,
one has
$$\eqalign{
\Sigma=& {1\over \sqrt 2}\b D_+ D_- V=\sigma - i \sqrt 2\theta^+
\b\lambda_+ -i\sqrt 2\b\theta{}^-\lambda_- +
\sqrt 2\theta^+\b\theta{}^-(D-iv_{01})\cr &-i\b\theta{}^-\theta^-(\partial_0
-\partial_1)\sigma-i\theta^+\b\theta{}^+(\partial_0+\partial_1)\sigma
+\sqrt 2\b\theta{}^-\theta^+\theta^-(\partial_0-\partial_1)\b\lambda_+
\cr &+\sqrt 2\theta^+\b\theta{}^-\b\theta{}^+(\partial_0+\partial_1)\lambda_-
-\theta^+\b\theta{}^-\theta^-\b\theta{}^+(\partial_0{}^2-\partial_1{}^2)\sigma.
\cr}\eqn\ucu$$
(The complicated terms involving derivatives are determined from the first
three terms by the twisted chiral condition $\b\cd_+\Sigma=\cd_-\Sigma=0$;
they could be eliminated
by the twisted version of $x^m\to y^m+i\theta^\alpha\sigma^m_{\alpha\dot\alpha}
\b\theta{}^{\dot\alpha}$.)

\subsection{Lagrangians}

It is straightforward to write Lagrangians in superspace.
We will consider only super-renormalizable theories that will
be particularly simple representatives of their universality classes.

These Lagrangians are of the form
$$L=L_{{\mit kin}}+L_{{\mit W}}+
L_{{\mit gauge}}+L_{{\mit D,\theta}}
\eqn\ommo$$
where the four terms are respectively the kinetic energy of chiral superfields,
the superpotential interaction, the kinetic energy of gauge fields, and the
Fayet-Iliopoulos term and theta angle, all
constructed as follows.

For simplicity, we write the formulas for the case of an abelian gauge group
$U(1)^s$, described as above by vector superfields $V_a, a=1\dots s$.
This will be the case of primary interest.
We assume that there are $k$ chiral superfields $\Phi_i$ of charges
$Q_{i,a}$.  The $\Phi_i$ can be interpreted as coordinates on a copy of
$Z={\bf C}^k$; and their kinetic energy is determined by a Kahler metric
on $Z$.  For superrenormalizability, this metric should be flat.
The Lagrangian corresponding to such a metric is, with a suitable
choice of coordinates,
$$L_{\mit kin}=\int\d^2 y\, \d^4\theta \sum_i\b\Phi_i\Phi_i
  =\int\d^2 y \d^4\theta \sum_i \b\Phi_{0,i} e^{2\sum_a Q_{i,a}V_a}
\Phi_{0,i}. \eqn\humpth$$
(The $\Phi_{0,i}$ are defined as in \convwrite\ by $\Phi_i=\exp(\sum_aQ_{i,a}
V_a)\Phi_{0,i}$.)  In components this becomes
$$\eqalign{L_{\mit kin}=&\sum_i\int\d^2y\left(-D_\rho\b \phi_i D^\rho \phi_i
+i\b\psi_{-,i}(D_0+D_1)\psi_{-,i}+i\b\psi_{+,i}(D_0-D_1)\psi_{+,i}+|F_i|^2
\right.\cr
& \left.-2\sum_a\b\sigma_a \sigma_a Q_{i,a}{}^2\b \phi_i \phi_i
-  \sqrt 2\sum_aQ_{i,a}\left(\b\sigma_a\b\psi_{+i}\psi_{-i}
+\sigma_a\b\psi_{-i}\psi_{+i}\right) + \sum_aD_a Q_{i,a}
\b \phi_i\phi_i\right.\cr &
\left. -\sum_ai\sqrt 2 Q_{i,a}\b \phi_i(\psi_{-,i}
\lambda_{+,a}-\psi_{+,i}\lambda_{-,a})
-\sum_ai\sqrt 2 Q_{i,a} \phi_i(\b \lambda_{-,a}
\b \psi_{+,i}-\b\lambda_{+,a}\b\psi_{-,i})\right)
\cr}\eqn\concor$$
where the world-sheet metric is $\d s^2=-(\d y^0)^2+(\d y^1)^2$.

The other part of the Lagrangian involving the chiral superfields is
constructed from a gauge invariant
holomorphic function $W$ on $Z$ known as the superpotential.
In our models, $W$ will always be a polynomial.
The corresponding part of the Lagrangian is
$$L_{\mit W}=-\int \d^2y \d\theta^+\d\theta^-\left. W(\Phi_i)\right|
_{\b\theta^+=\b\theta^-=0} -{\mit h.c.} \eqn\girg$$
This is integrated over only half of the odd coordinates, an operation
that is supersymmetric only because the $\Phi_i$ are chiral superfields
and $W$ is holomorphic.  In components,
$$L_{\mit W}=-\int\d^2y\left(F_i{\partial W\over \partial \phi_i}+
{\partial^2W\over\partial \phi_i\partial \phi_j}
\psi_{-,i}\psi_{+,j}\right)-h.c.
\eqn\urgo$$

The gauge kinetic energy is constructed from the twisted chiral superfields
$\Sigma_a$ defined as above.  Introducing gauge coupling constants
$e_a,\,\,a=1\dots s$, the gauge kinetic energy is
$$L_{\mit gauge}=-\sum_a{1\over 4e_a{}^2}
\int \d^2y \d^4\theta\; \b\Sigma_a \Sigma_a. \eqn\jucu$$
In components this is
$$\eqalign{L_{\mit gauge} = &
\sum_a{1\over e_a{}^2}\int \d^2y\left({1\over 2}v_{01,a}{}^2+{1\over 2}
D_a{}^2+i\b\lambda_{+,a}(\partial_0-\partial_1)\lambda_{+,a}
\right.\cr &~~~~~~~~
\left.+i\b\lambda_{-,a}(\partial_0+\partial_1)\lambda_{-,a}
-|\partial_\rho\sigma_a|^2\right).\cr} \eqn\ucuc$$

Certain additional terms involving the gauge multiplet
will play an essential role.
For simplicity I write the following formulas for the case of a single
gauge multiplet $V$, that is a gauge group that is just $U(1)$.  The
generalization is obtained simply by summing over the various components,
as in \ucuc.

By inspection of the supersymmetry transformation laws, one can see that
the Fayet-Iliopoulos term
$$-r \int\d^2 y~ D \eqn\hunc$$
is supersymmetric.  This is usually written in superspace as $-\int \d^2 y
\,\,\d^4\theta \,\,V$, but that is an unsatisfactory representation as $V$ is
not gauge invariant.  I will give a better version momentarily.
Another important coupling is the theta angle,
$${\theta\over 2\pi}\int \d v={\theta\over 2\pi}\int \d^2y \,\,\,v_{01}.
\eqn\ugu$$
$\theta$ should be regarded as an angular variable (the physics
is periodic in $\theta$ with period $2\pi$, but perhaps not smooth
or even continuous in $\theta$), since the quantity $\int \d v/2\pi$
measures the first Chern class of the abelian gauge field and is
always integral (with appropriate boundary conditions).

The above interactions can be neatly written in superspace using the
twisted chiral superfield $\Sigma$.
Indeed,
$$\eqalign{\int \d^2y\d\theta^+\d\b\theta^- \Sigma|_{\theta^-=\b\theta^+=0}
&=\sqrt 2\int\d^2 y\left(D-iv_{01}\right) \cr
\int \d^2y\d\theta^-\d\b\theta^+ \b\Sigma|_{\theta^+=\b\theta^-=0}
&=\sqrt 2\int\d^2 y\left(D+iv_{01}\right). \cr} \eqn\bubbb$$
{}From \bubbb\ it follows that we can write
$$\eqalign{L_{\mit D,\theta}=&
\int\d^2y\left(-rD+{\theta\over 2\pi}v_{01}\right)
\cr =& {it\over 2\sqrt 2}\int \d^2y\d\theta^+\d\b\theta^- \left.\Sigma\right|
_{\theta^-=\b\theta{}^+=0}
-{i\b t\over 2\sqrt 2}\int\d^2y\d\theta^-\d\b\theta^+
\left.\b\Sigma\right|_{\theta^+=\b\theta{}^-=0}\cr}
\eqn\repos$$
with
$$ t=ir+{\theta\over 2\pi}.      \eqn\hopos$$

\subsection{Twisted Chiral Superpotential}

\REF\dadda{A. D'Adda, A. C. Davis, P. DiVecchia, and P. Salomonson,
``An Effective Action For The $\CP^{n-1}$ Model,'' Nucl. Phys.
{\bf B222} (1983) 45.}
\REF\hubo{T. Hubsch, ``Of Marginal Kinetic Terms And Anomalies,''
Mod. Phys. Lett. {\bf A6} (1991) 1553}
\repos\ has a generalization that may be unfamiliar as it has no close analog
in four dimensions.  (See however [\dadda,\rocverl] for closely
related matters.)  We can introduce a holomorphic
``twisted superpotential'' $\widetilde W(\Sigma)$ and write
$$\Delta L= \int \d^2y\d\theta^+\d\b\theta^- \left.\widetilde W(\Sigma)
\right|_{\theta^-=\b\theta{}^+=0}+h.c.
        \eqn\opos$$
In components, this is
$$\Delta L=\int \d^2y\left(\sqrt 2\widetilde W'(\sigma)(D-iv_{01})+
2\widetilde W''(\sigma)\b\lambda_+\lambda_-\right)+h.c.      \eqn\lopos$$
This reduces to $L_{\mit D,\theta}$ precisely if $\widetilde W$ is the
linear function $\widetilde W(x)=itx/2\sqrt 2$.

In this paper, we will assume that the microscopic $\widetilde W$ function
is linear,
partly to ensure the $R$
invariance that is discussed presently.
Even so, we will find that a more elaborate twisted superpotential may
be generated by quantum corrections.
The following comments will be useful background for that discussion.

Consider the twisted superpotential $\widetilde W(x)=-x\ln x\cdot p/\sqrt 2$
($p$ being a constant).  We compute
$$\Delta L= -p\int \d^2y \left((\ln \sigma +1)(D-iv_{01})+\sqrt 2{1\over\sigma}
\b\lambda_+\lambda_-\right)-h.c. \eqn\hondo$$
Because $\ln\sigma$ is only well-defined modulo $2\pi i\Z$, \hondo\
is not well-defined as a real-valued functional.  However, in quantum
mechanics it is good enough if $\Delta L$ is well-defined modulo $2\pi$.
Since $\int \d^2y\,\,
v_{01}$ takes values in $2\pi \Z$, $\Delta L$ is well-defined
modulo $2\pi$ if and only if
$$ 4\pi p\in \Z.      \eqn\ubu$$

Because \hondo\ has a singularity at $\sigma=0$, its possible generation
by quantum corrections would occur, if at all, only
in an approximation whose validity would be limited to large $\sigma$.
This is precisely what we will find.
For future use, let us note that if $\sigma$ is large and slowly varying,
then the addition of \hondo\ to the Lagrangian would have the effect
of shifting $r$ to an effective value
$$r_{\mit eff} = r + 2p\ln |\sigma|. \eqn\zubu$$

\subsection{Symmetries}

The models that we have constructed have $N=2$ supersymmetry in two
dimensions, that is two left-moving supersymmetries (acting on $\theta^-,
\b\theta{}^-$)
and two right-moving ones (acting on $\theta^+,\b\theta{}^+$).
It is natural to ask if the models also have left and right-moving $R$
symmetries.  A right-moving $R$ symmetry is a $U(1)$ symmetry under
which $\theta^+\to e^{i\alpha}\theta^+,\,\,\b\theta{}^+\to e^{-i\alpha}\b
\theta{}^+$, while $\theta^-,\b\theta{}^-$ are invariant.  A left-moving
$R$ symmetry obeys the analogous condition with $+$ and $-$ exchanged.

Left and right-moving $R$ symmetries are important because they are part
of the $N=2$ superconformal algebra.  If present, they give important
information about possible conformally invariant limits
of the models under discussion.

The existence of such left and right-moving symmetries in the models
under discussion is fairly
natural from the four dimensional point of view.  Many four dimensional
supersymmetric theories would have a single $R$ symmetry, with charges
$+1$ for $\theta^+$ and $\theta^-$, and $-1$ for $\b\theta^\pm$.
In addition, in the process of dimensional reduction to two dimensions,
we potentially get a second $U(1)$ symmetry -- corresponding to rotations
of the extra dimensions.  Under this symmetry, $\theta^+$ and $\theta^-$
transform oppositely.  By taking suitable combinations of these operations,
one may hope to find the left or right-moving $R$ symmetry, under which
$\theta^+$ or $\theta^-$ should be invariant.

It is easy to find these symmetries explicitly in the above two dimensional
formulas.  First we suppress the superpotential interaction $L_{\mit W}$,
and we also assume that any twisted superpotential interaction, if present,
is a linear term of the specific form $L_{\mit D,\theta}$.
Under these assumptions, the left and right-moving $R$ symmetries
are easy to find explicitly.  For instance, there is a right-moving
$R$ symmetry ($\epsilon^+=-\epsilon_-$ has charge 1, $\epsilon^-=\epsilon_+$
is invariant) under which the non-zero charge assignments are
as follows: $(\psi_{+i},F_i,\sigma_a,\lambda_{-a})$ have charges
$(-1,-1,1,1)$; their complex conjugates have opposite charge; and other
fields are invariant.  A left-moving $R$ symmetry ($\epsilon^-=\epsilon_+$ has
charge 1, $\epsilon^+=-
\epsilon_-$ is invariant) likewise can be constructed
under which $(\psi_{-i},F_i,\sigma_a,\lambda_{+a})$ have charges
$(-1,-1,-1,1)$ and other fields are neutral.
Let us call the charges generating the right- and left-moving symmetries
$J_R$ and $J_L$.

One of the most important properties is that $J_R$ and $J_L$ may be anomalous.
The reason for this is that $J_R$, for instance, couples to right-moving
charged fermions $\psi_{+i},\b\psi_{+i}$, but the left-moving fermions
that it couples to are neutral.  The condition that there should be no
gauge anomalies that would spoil conservation of $J_R$ is that
$$\sum_i Q_{i,a}=0,\;\;\;\;\;{\rm for}\;a=1\dots s.\eqn\gogo$$
The same condition is needed to ensure that $J_L$ is actually a valid
symmetry.  (In any event, regardless of the $Q_{i,a}$, the sum $J_L+J_R$
is conserved.)
We will see later that various other good things happen when \gogo\ is
imposed.

Before going on, it is important to note that the models under discussion,
if we ignore the superpotential, have additional (non-anomalous) symmetries
that commute with both left and right-moving supersymmetries and therefore
could be added to the $J_L$ and $J_R$ introduced above.  These are global
transformations of the chiral superfields of the form
$$\Phi_i\to \exp(i\alpha k_i)\Phi_i, \eqn\nucu$$
with arbitrary $k_i$; they commute with supersymmetry since a common
transformation is assumed for each component of $\Phi_i$.

Are these $R$ symmetries still valid in the presence of a superpotential?
The couplings coming from $\int \d^2\theta \,W$ have charge $1$ under
$J_R$ as we have defined it hitherto.  To save
the situation, we must add to $J_R$
a transformation of the form \nucu\ under which $W\to e^{-i\alpha}W$.
The superpotential $W$ is said to be quasi-homogeneous if $k_i$ exist
so that $W$ transforms as indicated.  Thus, by adding additional terms
to the original $J_R$ (and analogous additional terms
for $J_L$), one can find left- and right-moving $R$
symmetries precisely in case \gogo\ holds and $W$ is quasi-homogeneous.

The twisted superpotential violates $R$-invariance unless it takes
the linear form that leads to what we have called $L_{\mit D,\theta}$.
Therefore, in the $R$-invariant case, the quantum corrections cannot
generate a more complicated twisted chiral superpotential.

\chapter{THE C-Y/L-G CORRESPONDENCE}

We now turn to our problem of shedding some light on the correspondence
between Calabi-Yau and Landau-Ginzburg models.  In doing so, we will
consider only the simplest situation, with a gauge group that is simply
$G=U(1)$, and thus a single vector superfield $V$.
The generalizations will be relegated to \S4.

Happily, of all the formulas of \S2, we primarily need
only a few of the simpler ones.
The fields $D$ and $F_i$ enter in the Lagrangian \ommo\ as
``auxiliary fields,'' without kinetic energy.  One can solve for them
by their equations of motion to get
$$\eqalign{ D & =-e^2\left( \sum_i Q_i|\phi_i|^2 -r\right) \cr
F_i &= {\partial W\over \partial \phi_i}.\cr}   \eqn\umbo$$
The potential energy for the dynamical scalar fields $\phi_i,\sigma$
can then be written
$$U(\phi_i,\sigma)={1\over 2e^2}
D^2+\sum_i|F_i|^2  +2\b\sigma \sigma\sum_iQ_i{}^2|\phi_i|^2.
\eqn\bumbo$$

\section{The Model}

Now, let me explain the situation on which we will focus.
We take the chiral superfields $\Phi_i$ to be $n$ fields $S_i$ of charge 1,
and one field $P$ of charge $-n$.
This obviously ensures condition \gogo\
for anomaly cancellation and $R$ invariance.
Lower case
letters $s_i$ and $p$ will denote the bosonic fields in the supermultiplets
$S_i$ and $P$.
We take the superpotential to be the gauge invariant function
$$W = P \cdot G(S_1,\dots,S_n) , \eqn\uccu$$
with $G$ a homogeneous polynomial of degree $n$.
Notice that this $W$ is quasi-homogeneous; one can
pick the $k_i$ in \nucu\ to be $-1$ for $P$ and 0 for $S_i$
or (by adding a multiple of the gauge generator) 0 for $P$ and $-1/n$ for
$S_i$.  The superpotentials we consider later in this paper will always
be quasi-homogeneous in a similar obvious way; this will not be spelled
out in detail in subsequent examples.

We want to choose $G$ to be
``transverse'' in the sense that the equations
$$ 0={\partial G\over \partial S_1}=\dots={\partial G\over\partial S_n}
    \eqn\uvuv$$
have no common root except at $S_i=0$.
This property ensures that if we think of the $S_i$ as homogeneous
variables, then the hypersurface $X$ in $\CP^{n-1}$ defined by the equation
$G=0$ is smooth.  A generic homogeneous polynomial
has this property.  An analogous transversality condition enters
in each of our later models.

For the model just introduced, the bosonic potential \bumbo\ is
$$U=|G(s_i)|^2 +|p|^2\sum_i\left|{\partial G\over\partial s_i}\right|^2+
{1\over 2e^2}D^2+
2|\sigma|^2\left(\sum_i|s_i|^2
+n^2|p|^2\right) \eqn\vuvu$$
with
$$D=   -e^2\left(\sum_i \b s_i s_i-n \b p p -r\right).  \eqn\occoc$$

Now, let us discuss the low energy physics for various values of $r$.
First, we take $r>>0$.  In this case, obtaining $D=0$ requires that
the $s_i$ cannot all vanish.  That being so, vanishing of the
term $|p|^2\sum_i|\partial_iG|^2$ in the potential requires
(in view of transversality of $G$)
that $p=0$. Vanishing of $D$ therefore gives us precisely
$$ \sum_i\b s_i s_i = r.          \eqn\icco$$
Dividing the space of solutions of \icco\ by the gauge group $U(1)$,
we get precisely a copy of complex projective space $\CP^{n-1}$,
with Kahler class proportional to $r$.  Finally,
we must set $G=0$ to ensure the vanishing
of the first term in the formula \vuvu\ for the potential, and $\sigma=0$
to ensure the vanishing of the last term.

So the space of classical vacua is isomorphic to the hypersurface
$X\subset \CP^{n-1}$ defined by $G=0$.  It is a well-known fact
that a smooth hypersurface of degree $k$ in $\CP^{n-1}$ is a Calabi-Yau
manifold if and only if $k=n$.  We have not accidentally
stumbled upon the Calabi-Yau condition.  We picked $P$ to be of charge
$-n$, and hence $G$ of degree $n$, to ensure the condition \gogo\ for
anomaly-free $R$-invariance.  Anomaly free $R$-invariance of the underlying
model ensures such invariance of the effective low energy sigma model;
but for sigma models, anomaly free $R$-invariance is equivalent to the
Calabi-Yau condition.

All modes
other than oscillations tangent to $X$ have masses at tree level.
(The gauge field $v$ gets a mass from the Higgs mechanism; masses for $p$,
$\sigma$, and modes of $s_i$ not tangent to $X$ are visible in \uvuv.)
The low energy theory therefore is a sigma model with target space $X$
and Kahler class proportional to $r$.

Now we consider the case of $r<<0$.  In this case, the vanishing
of $D$ requires $p\not = 0$.  The vanishing of $|p|^2\sum_i|\partial_iG|^2$
then requires (given transversality of $G$) that all $s_i=0$.
This being so, the modulus of $p$ must in fact be $|p|=\sqrt{-r/n}$.
By a gauge transformation, one can fix the argument of $p$ to vanish.
So the theory has a unique classical vacuum, up to gauge transformation.
In expanding around this vacuum, the $s_i$
are massless (for $n\geq 3$).
These massless fields are governed by an effective
superpotential that can be determined by integrating out the massive
field $p$; integrating out $p$ simply means in this case setting $p$
to its expectation value.  So the effective superpotential of the low
energy theory is $\widetilde W = \sqrt{-r}\cdot W(s_i)$.
The factor of $\sqrt{-r}$ is inessential, as it can be absorbed in rescaling
the $s_i$.
This effective superpotential
has (for $n\geq 3$ ) a degenerate critical point at the origin, where
it vanishes up to $n^{th}$ order.  A theory with a unique
classical vacuum state governed by a superpotential with a degenerate
critical point is usually called a Landau-Ginzburg theory.

In this case, we actually get a Landau-Ginzburg orbifold, that is an
orbifold of a Landau-Ginzburg theory, for the following reason.
The vacuum expectation value of $p$ does not completely destroy
the gauge invariance; rather, it breaks $U(1)$ down to a $\Z_n$
subgroup $E$ that acts by $s_i\to \zeta s_i$, where $\zeta $ is an $n^{th}$
root of 1.  This residual gauge invariance means that what we get is
actually a $\Z_n$ orbifold of a Landau-Ginzburg theory.
This follows from
a general and elementary though perhaps not universally recognized fact;
orbifolds (as the term is usually used in quantum field theory) are equivalent
to theories with a finite gauge group.

In fact,
in calculating the path integral on a Riemann surface $\Sigma$,
instead of expanding about the absolute minimum of the action
at $p=\sqrt{- r}$, $s_i=v=0$, and gauge transformations thereof,
we can expand around configurations that are gauge equivalent
to $p=\sqrt{-r}$, $s_i=v=0$ only locally.
In such configurations, the gauge fields $v$ may have monodromies.
As the monodromies must leave $p$ invariant, they take values in the group
$E$ of $n^{th}$ roots of unity.
The construction of the low energy approximation
to the path integral involves a sum over the possible
$E$-valued monodromies.  But this sum over monodromies
is precisely the usual operation
of summing over $E$-twists in all channels
by which one constructs the quantum field theory of the $E$-orbifold.
Thus, in general, orbifolds can be regarded as a special case
of gauge theories in which the gauge group is a finite group.

Of course, both for $r>>0$ and for $r<<0$, the integration
over the massive fields to get an effective theory for the massless
fields cannot just be done classically; we must consider quantum
corrections to the sigma model or Landau-Ginzburg orbifold obtained
above.  We will look at this more closely in \S3.2; suffice it to say
here that in the $R$-invariant case, for sufficiently big $|r|$,
integrating out the massive fields just corrects the parameters
in the effective low energy theory without a qualitative change
in the nature of the system.

\subsection{The Moral Of The Story}

The above construction suggests that, rather than Landau-Ginzburg
being ``equivalent'' to Calabi-Yau, they are two different phases
of the same system.  In fact, from this point of view Landau-Ginzburg
looks like the analytic continuation of Calabi-Yau to negative
Kahler class.  We will reexamine the moral
of our story after some further analysis.

\section{The ``Singularity''}

To extract any precise conclusions about the relation between
the Landau-Ginzburg and Calabi-Yau models, we will have to deal with
the apparent singularity at $r=0$ that separates them.

The problem is not that there might be a phase transition at $r=0$.
In our applications, we are interested either in quantizing the theory
on a circle (compact, finite volume), or on performing path integrals
on a compact Riemann surface.  Either way,
we will not see the usual singularities associated
with phase transitions; those depend crucially on having infinite
world-sheet volume.  The only singularities we might see would
be due to failure of effective compactness of the target space.

For instance, if the $\sigma$ field were absent (we will see in \S6 that
it can be suppressed preserving $(0,2)$ supersymmetry), there could
be no singularity at $r=0$ in finite world-sheet volume.  For without
the $\sigma$ field, the only possible zero of the classical bosonic
potential at $r=0$ would be at $s_i=p=0$.
Moreover, the locus in which the classical potential is less than
any given number $T$ would likewise be compact.
These conditions mean that in quantization on a circle,
quantum states of any given energy decay exponentially at large
values of the fields, so that the spectrum is discrete.
Such a discrete spectrum varies continuously
-- though perhaps in a difficult-to-calculate fashion -- with the
parameters.

The $\sigma$ field changes the picture even for $r\not = 0$.
Though the space of classical vacua is compact for any $r$, and was
determined above, the space on which the classical potential is
less than some value $T$ is compact only for $T$ small enough.
In fact, for $s_i=p=0$, the classical potential has the constant value
$D^2/2e^2=e^2r^2/2$, independent of $\sigma$.  This means that only the
low-lying part of the spectrum -- at energies such that the region
at infinity is inaccessible -- is discrete.  If the theory is quantized
on a circle of radius $R$, then at energies above some critical $T$,
the theory has a continuous spectrum, coming from wave-functions supported
near $s_i=p=0$, $|\sigma|>>0$.  Semi-classically, the critical value
of $T$ appears to be
$$T_{\mit cr}={e^2r^2\over 2}\cdot 2\pi R. \eqn\moco$$

Before going on, let us assess the reliability of this estimate.  The
non-zero value of $T$ came from the expectation value of the $D$
field in the region $s_i,p$ near zero, $|\sigma|>>0$.
In that region, the chiral superfields, including
$s_i$, $p$ and their superpartners, have masses
proportional to $|\sigma|$; the gauge multiplet ($\sigma$, $v$, and $\lambda$)
is massless.  If one simply sets $s_i$ and $p$
to zero, the gauge multiplet is described by a free field theory
and (except for a subtlety about the $\theta$ angle that we come to later)
there are no quantum corrections at all
to the classical value \moco.  Naively, integrating out fields with
masses proportional to $|\sigma|$ will give only corrections proportional
to negative powers of $|\sigma|$ and should not change that conclusion.
This will be so if the ultraviolet behavior is good enough.

To see what really happens, consider a general
model with chiral superfields $\Phi_i$ of charge $Q_i$.  The
one loop correction to the vacuum expectation value of $-D/e^2=\sum_iQ_i
|\phi_i|^2-r$ will be
$$\delta\left\langle -{ D\over e^2}\right\rangle
 = \sum_iQ_i \int {\d^2k\over (2\pi)^2}{1\over k^2+2|\sigma|^2+\dots}
. \eqn\jjj$$
(The $\dots$ are $\sigma$-independent contributions to the masses.)
This vanishes for large $\sigma$ if
$$\sum_i Q_i = 0.      \eqn\colpo$$
If \colpo\ is not obeyed, then \jjj\
diverges.  To renormalize the theory, we subtract the value of $\delta D$
at, say, $|\sigma|=\mu$.   The subtraction is interpreted as a redefinition
of the parameter $r$.  After the subtraction, one has for the large
$\sigma$ behavior
$$\eqalign{
\delta \left\langle -{D\over e^2}\right\rangle
 & =\sum_iQ_i\int{\d^2 k\over (2\pi)^2}\left({1\over k^2+2|\sigma|^2}
 -{1\over k^2+2\mu^2}\right)\cr
& =\sum_iQ_i\cdot \left(\ln(\mu/|\sigma|)\over 2\pi\right)  .\cr}
\eqn\ggg$$
This formula diverges for $|\sigma|\to \infty$.
(The formula also diverges for $\sigma$ near 0, but this effect
is misleading; for $\sigma$ near 0, the point $s_i=p=0$ about which
we expanded is unstable and the approximations have been inappropriate.)
The divergence
means that quantum corrections are large in this region of field space
if $\sum_iQ_i\not=0$,
and crucial properties cannot be read off from the classical Lagrangian.  The
correction to $D$ can be interpreted in terms of a
$\sigma$-dependent effective value of $r$,
$$r_{\mit eff}\sim r+{1\over 2\pi}\sum_iQ_i\cdot \ln(|\sigma|/\mu).\eqn\boblo$$

One may wonder what superspace interaction can have this $\sigma$ dependent
effective value of $r$ as one of its consequences.  Looking back
to equation \zubu, the answer is clear.  We have generated
a twisted chiral superpotential, of precisely the form considered
in \hondo, with
$$4\pi p =\sum_iQ_i \eqn\jupu$$
-- in agreement with our general observation that $4\pi p$ must be an integer.
(In the case of the $\CP^{n-1}$ model, to which we will specialize later,
this effect was first computed from a different viewpoint
by D'Adda et. al. [\dadda].  The general formula has also been
conjectured [\hubo].)
I leave it to the curious reader to verify, by further computations
in the large $\sigma$ region, the $\sigma$-dependent effective value of
$\theta$
that is also predicted by this twisted chiral superpotential.

Before discussing the consequences of this computation, let us
assess its relevance.
Is the specific quantum correction that
we just evaluated the only one that matters?  In fact,
for large $\sigma$, this theory is a theory of massless free fields weakly
coupled by superrenormalizable interactions to fields of mass proportional
to $\sigma$.  Perturbation theory will be uniformly valid
(and in fact increasingly good) for large $\sigma$, unless there
are ultraviolet divergences.  (Such divergences can be translated upon
renormalization into effects that grow for large $\sigma$, as we have just
seen.)  Because of the superrenormalizability  of these models,
and the non-renormalization theorems of the superpotential, the only
divergent quantum correction is the one-loop renormalization
of $r$ that we have just encountered.
(A roughly analogous fact in four dimensional
renormalizable supersymmetric theories is that the only quadratic divergence is
the one loop renormalization of $r$, again proportional
to $\sum_iQ_i$.)  That is why this is the important quantum
correction.

\REF\coleman{S. Coleman, ``More On The Massive Schwinger Model,''
Ann. Phys. (NY) {\bf 101} (1976) 239. }

\subsection{The Case $\sum_iQ_i=0$}

Let us now discuss the consequences of the above computation.
First we assume that $\sum_iQ_i=0$.  In that case, for large $\sigma$
we can simply throw away the chiral superfields, with their masses
of order $|\sigma|$.
We still have to look at the effective theory of the massless
gauge multiplet; this multiplet has the effective
Lagrangian $L_{{\mit gauge}}+
L_{D,\theta}$.  This effective theory is a free theory, and supersymmetry
ensures that the usual quantum corrections to the ground state energy
from zero point fluctuations in $\sigma$ and $\lambda$ cancel.
But for the gauge field $v$, an important subtlety arises.
The Lagrangian for $v$ is
$$L=\int_\Sigma \d^2y\,\left(
{1\over 2 e^2}v_{01}^2+{\theta\over 2\pi}v_{01}\right).\eqn\ripo$$
The significance of the $\theta$ term in abelian gauge theory in
two dimensions was explained long ago [\coleman].  It induces a constant
electric field in the vacuum, equal to the electric field
due to a charge of strength $\theta/2\pi$.  This constant electric
field contributes to the energy of the vacuum.
This contribution has the value that one might guess by simply
minimizing the Lagrangian with respect to $v_{01}$, namely
$(e^2/2)\cdot (\theta/2\pi)^2$.
More exactly, this is so
as long as $|\theta|\leq \pi$.  Otherwise, by a process involving
pair creation, the contribution to the energy can be reduced to
the minimum value of $(e^2/2)\cdot ((\theta/2\pi) -n)^2$ for $n\in \Z$.
The contribution of the theta term
to the vacuum energy density is thus
in general
$${e^2\over 2}\left(\widetilde \theta\over 2\pi\right)^2 \eqn\poxx$$
where $|\widetilde \theta|\leq \pi$ and $\widetilde\theta-\theta
\in 2\pi \Z$.
In particular the vacuum energy density is a continuous and
periodic but not smooth function
of $\theta$ with period $2\pi$.

We thus can determine the {\it exact},  quantum-corrected analog
of \moco: the minimum energy density of a quantum state at large $\sigma$ is
$$U={e^2\over 2}\left(r^2+\left({\widetilde\theta\over 2\pi}\right)^2\right).
\eqn\minen$$
The minimum energy of a quantum state obtained in quantizing the
gauge multiplet
on a circle of circumference $2\pi R$ is
$$T'_{\mit cr}=U\cdot 2\pi R. \eqn\moco$$
The quantum energy spectrum is discrete and continuously varying
for energies below $T'_{\mit cr}$, for at such energies the target
space is effectively compact, the region of large $\sigma$ being
inaccessible.
As long as either $r$ or $\theta$ is non-zero, $T'_{\mit cr}$ is non-zero,
and there is a range of energies with such a discrete spectrum.

Another statement along a similar lines is important in
some of our applications.
Instead of considering the energy $H$, one could consider
$H+P$, $P$ being the momentum.  Since the vacuum has $P=0$ and particle
excitations contribute a non-negative amount to $H+P$, the conclusion
is the same: the spectrum is discrete for $H+P<T'_{\mit cr}$.

\REF\witmir{E. Witten, ``Mirror Manifolds And Topological Field
Theory,'' in S.-T. Yau, ed., {\it Essays On Mirror Manifolds}
(International Press, 1992).}
In all of our applications, when we try to actually learn something
from the relation between Landau-Ginzburg and
Calabi-Yau models, we will use some sort of generalized topological
reasoning.   In the simplest case, we consider the index of
one of the supercharges (equal to $\Tr(-1)^F$ or the elliptic genus,
as explained below).
Given a discrete and continuously varying spectrum for $H$ (or $H+P$)
below some positive constant, these indices can be computed by counting
the states of $H=0$ (or $H+P=0$), and in particular
are well-defined and invariant
under deformation.  Therefore, these quantities
are independent of $r$ and $\theta$ except at $r=\theta=0$, and
we can interpolate safely from $r>0$ to $r<0$ by
taking $\theta\not= 0 $.

We will also consider more delicate quantities involving couplings
of states of $H=0$ (or $H+P=0$).
Some of these quantities, related to Yukawa couplings, for instance,
are predicted to be independent of $r,\theta$,
while others are predicted to vary holomorphically in $t$.
These predictions depend on arguments
(see [\witmir], for instance) that rest ultimately on integration
by parts in field space.  These arguments might fail if the integrand
of the path integral were to behave badly for $\sigma\to \infty$.
However, as long as $T'_{\mit cr}\not= 0$, wave functionals
of states  of $H=0$ (or $H+P=0$) vanish exponentially for $\sigma\to\infty$.
Path integral representations of couplings of such states are then
highly convergent, and integration by parts in field space is justified.

In sum, then, the only real singularity, for the topological quantitities
in finite volume, is at $r=\theta=0$.
By taking $\theta\not= 0$, we can continue happily from Calabi-Yau
to Landau-Ginzburg.

\subsection{What Happens If $\sum_iQ_i\not= 0$}

Now we want to analyze what happens if $\sum_iQ_i\not=0$.

In our classical Lagrangian, the
twisted superpotential for the gauge multiplet $\Sigma$
was simply a linear function
$\widetilde W(\Sigma)=it\Sigma/2\sqrt 2$.
One might suspect that just as an ordinary
$W$ leads to a bosonic potential $\sum_i|\partial_iW|^2$, a twisted
superpotential would lead to a potential energy
$$\left|{\partial \widetilde W(\sigma)\over\partial\sigma}\right|^2.\eqn\gug$$
This is almost true, except for a factor that comes from the normalization
we used for the gauge multiplet.  Indeed,
the potential energy in \minen\ is $U=4e^2|\widetilde W_Q{}'|^2$ with
$\widetilde W_Q(x)=i\widehat tx/2\sqrt 2$; here $\widehat t = t+n$,
with $n\in \Z$ chosen to minimize $U$.
Notice that the mechanism for appearance of this (almost)
$|\widetilde W'|^2$ term in the energy
is rather subtle; it arises partly from integrating out the $D$
auxiliary field and partly from the effects of the $\theta$ angle.

The formula \gug\ is still valid, for the same reasons (except for the same
modifications), for computing the potential energy induced by
the more general twisted superpotential
$$\widetilde W = {1\over \sqrt 2}
        \left({it \Sigma\over 2}
 -{\sum_i Q_i\over 2\pi}\Sigma\ln(\Sigma/\mu)
            \right)\eqn\mucu$$
that incorporates the quantum correction.  After all, for large $\sigma$
we simply have a theory with a weak $\sigma$ dependence in the
effective values of $r$ and $\theta$,
and we can compute the dependence of the energy
on $r$ and $\theta$ just as we did when these were strictly constant.
So the effective potential for $\sigma$ is just
$$U(\sigma)={e^2\over 2}\left|{i\widehat t}-\sum_i
{Q_i\over 2\pi}\left(\ln(\sigma/\mu)+1
\right)\right|^2.      \eqn\mico$$
Again $\widehat t = t + n$, where $n\in \Z$ is to be chosen
to minimize $U(\sigma)$.  $U(\sigma)$ is therefore
a continuous but not smooth function of $\sigma$ and $t$.

Now we can see that our analysis of the correspondence between sigma models
and Landau-Ginzburg models must be modified when $\sum_iQ_i\not= 0$
because the semi-classical determination of the vacuum structure does
not apply.  For $\sum_iQ_i>0$, there are new vacuum states, not predicted
in the semi-classical reasoning, for $r<<0$; and for $\sum_iQ_i<0$, there
are such new states for $r>>0$.  These new ground states are determined
by
$$ {i\widehat t}-{\sum_iQ_i\over 2\pi}\left(\ln(\sigma/\mu)+1\right)=0
\eqn\nucuu$$
and there are precisely
$|\sum_iQ_i|$ of them.
The new solutions are explicitly
$$\sigma={\mu\over e}\exp\left({2\pi i \widehat t\over \sum_iQ_i}\right);
\eqn\jdjdsx$$
there are $|\sum_iQ_i|$ of them because of the freedom $\widehat t\to
\widehat t+n$.
Of course, the only relevant solutions of \nucuu\
are those that arise for large $\sigma$, where the approximations are
valid; this is why we need $r$ large and of the appropriate sign.

\mico\ grows at infinity like $|\ln\sigma|^2$.  This ensures, in finite
world-sheet volume, that the model has no singularity at any $r$ or
$\theta$, as long as $\sum_iQ_i\not=0$.\foot{This is in accord with the
fact that for the $A$ model or the half-twisted model, introduced
in \S3.3, the instanton sums are always finite when $\sum_iQ_i\not=0$
because the anomalous $R$ symmetry ensures that any given amplitude
receives contributions only from finitely many values of the instanton
numbers. When $\sum_iQ_i=0$, the instanton sums are infinite sums
that can have singularities; we will see an example in \S5.5.}
Thus, a generalization of the sigma model/L.G. correspondence, taking
account the extra states at infinity, holds even when
$\sum_iQ_i\not=0$.

For illustration, let us see how the new states enter in the $\CP^{n-1}$
model.
We consider a model with gauge group $U(1)$ and $n$ chiral superfields
$S_i$, all of charge 1.  Gauge invariance requires the
superpotential to be $W=0$.
The classical potential is simply
$$U(s_i,\sigma)={e^2\over 2}\left(\sum_i\b s_is_i-r\right)^2
+2|\sigma|^2\sum_i|s_i|^2. \eqn\hogo$$
For $r>>0$, a classical vacuum must have $\sigma=0$, $\sum_i|s_i|^2=r$.
The space of such vacua, up to gauge transformation, is a copy of
$\CP^{n-1}$ with Kahler class proportional to $r$,
and the model at low energies is a sigma model with this
target space.  For $r<<0$, it appears on the other hand that supersymmetry
is spontaneously broken.  This poses a problem.
For $r>>0$, the supersymmetric index $\Tr(-1)^F$
of the theory\foot{Its definition is well-known and is given below.}
is equal to $n$ (the Euler characteristic of
$\CP^{n-1}$), while spontaneous supersymmetry
breaking would mean that the index
would be 0 for $r<<0$.
Given the $|\ln \sigma|^2$ behavior of the potential at infinity,
vacuum states cannot flow to or from infinity, and the index should be
constant.
To solve the problem, we need only
note that $\sum_iQ_i=n$.  So the expected $n$ zero energy states
are found by setting $s_i=0$ and $\sigma$ to a solution of \nucuu.

\REF\vafcec{C. Vafa and S. Cecotti, ``Exact Results For Supersymmetric
Sigma Models,'' Phys. Rev. Lett. {\bf 68} (1992) 903.}
Another continuation
of the $\CP^{n-1}$ sigma model to negative $r$ was discussed recently
[\vafcec].

\subsection{The Moral, Revisited}

Our arguments have been sensitive only to singularities crude enough
as to be visible even in finite world-sheet volume.
What is the phase structure in infinite world-sheet volume?
Are Calabi-Yau and Landau-Ginzburg separated by a true phase
transition, at or near $r=0$?

There is no reason that the answer to this question has to be
``universal,'' that is, independent of the path one follows
in interpolating from Calabi-Yau to Landau-Ginzburg
in a multiparameter space of not necessarily conformally invariant
theories.  Along a suitable path, there may well be a sharply defined
phase transition, while along another path there might not be one.
This seems quite plausible.

The situation would then be similar to the relation
between the gas and liquid ``phases'' of a fluid.
We would have one system which in one limit is well described as Calabi-Yau;
in another limit it is well described as Landau-Ginzburg.
But these would be two different limits of the same system, with no
order parameter or inescapable singularity separating them; and so
one might say (as is customary) that Landau-Ginzburg and Calabi-Yau are
``equivalent.''

It is believed that the theories we are studying all flow in the infrared
to conformal field theories.  If so, a particularly interesting type
of interpolation from Landau-Ginzburg to Calabi-Yau would be via
conformally invariant theories.  Is there a continuous family of
conformal field theories interpolating from Landau-Ginzburg to Calabi-Yau?
This is the only way that Landau-Ginzburg and Calabi-Yau could truly
be ``equivalent'' as conformal field theories.
In the case of topology change (which we will see to be analogous
in \S5.5), mirror symmetry gives
[\amg] a direct indication that the answer is ``Yes.''
In the present context, we can attempt to argue (not rigorously)
as follows.  First of all, any singularity or discontinuity of a family
of conformal field theories that holds on the real line must already
hold on the circle, since conformally the real line is just the circle
with a point deleted.  Now, in the above,
for linear sigma models we identified all of the singularities
that occur in finite volume: the only such singularity is
at $r=\theta=0$.  The key point to avoid a singularity
was the non-vanishing of the large $\sigma$
minimum energy density $U$, given in \minen.  This was an exact
formula, not depending on small $e$.
To approach the infrared limit, we should scale $e\to infinity$,
and it would appear that $U$ only grows in this limit, so that new
singularities would not arise.
This argument encourages me to believe that the singularities
we found for linear sigma models in finite volume are the only
ones that will appear in conformal field theory.
The potential trouble with the argument
is that the definition of $U$ involved a large $\sigma$
limit which might not commute with the large $e$ limit.

In any event,
semi-classically, just like liquid and gas, Calabi-Yau
and Landau-Ginzburg look like different ``phases,'' and I will use
that convenient, informal language.

\section{Applications}

Whether $r$ is positive or negative, the linear sigma models
investigated here are super-renormalizable rather than conformally invariant.
Even classically, to reduce the $r>>0$
theory to a Calabi-Yau nonlinear sigma model,  or to reduce the $r<<0$ theory
to a Landau-Ginzburg model, we have to take the limit
in which $e$ (the gauge coupling) goes to infinity.

But our arguments depended on having $e<\infty$.
Moreover, $e<<1$ is really the region in which the model is under
effective control, the semi-classical arguments being reliable.

If we do take $e\to\infty$ with $r>>0$, we will get a Calabi-Yau
non-linear sigma model with target a hypersurface $X$.  The Kahler metric $g$
on $X$ (along
with the corresponding Kahler form $\omega$) is induced
from the embedding in $\CP^{n-1}$ (with its standard metric).  It is known
that this metric on $X$ does {\it not} give a conformally invariant
model.  It is believed that there is (for large enough $r$) a unique
Kahler metric $g'$ on $X$, with a Kahler form $\omega'$ cohomologous
to $\omega$, that gives a conformally invariant sigma model.
The two models with metric $g$ and $g'$ differ
by a superspace interaction of the form
$$\int \d^2x\d^4\theta \,\,\,T,\eqn\tyr$$
where the function $T$ obeys  $\omega-\omega'=-i\partial\b\partial T$.

Consequently, even if we set $e=\infty$, the linear sigma model
considered here is not expected to coincide with
the conformally invariant nonlinear sigma
model.  They differ by the term \tyr, and moreover the function
$T$ is unknown.  So how can we learn anything relevant to the conformally
invariant model?

One answer is that the $T$ coupling, like the effects of finite $e$,
are believed to be ``irrelevant'' in the technical sense of the
renormalization group; that is, these effects are expected to
disappear in the infrared.  This is of theoretical importance
-- for instance it enters in heuristic discussions of the
Landau-Ginzburg/Calabi-Yau correspondence -- but difficult to
use as a basis for calculation.

A more practical answer depends on the fact that
$N=2$ theories in two dimensions are richly endowed with quantities
that are invariant under adding to the Lagrangian a term of the
form $\int \d^4\theta\left(\dots\right)$ (with $\dots $ a {\it gauge invariant}
operator).  Such quantities include $\Tr (-1)^F$, the elliptic
genus, and the observables of two twisted topological field theories
(the $A$ model and the $B$ model), and of a so far little studied
hybrid, the half-twisted model.  Let us call these the
quasi-topological quantities.  Any such quantities are invariant
under the addition of a term such as \tyr\ to the Lagrangian,
so they can be computed using the ``wrong'' Kahler metric on $X$.

Moreover, once one is reconciled to studying primarily
the quasi-topological quantities, it is easy to see that our problem
with the $e$
dependence can be
resolved in essentially the same way.     The $e$ dependent
term in the Lagrangian is
$$-{1\over 4e^2}\int \d^2y\d^4\theta   \,\,\b\Sigma\Sigma,\eqn\peff$$
so the quasi-topological quantitities are
independent of $e$.
The dependence on the Fayet-Iliopoulos and theta terms can be similarly
analyzed, with the result that some quasi-topological terms are invariant
under these couplings and others vary holomorphically in $t$.

\subsection{Indices}

The simplest quasi-topological quantities are simply the
indices of various operators (in canonical quantization on a circle).
Let $H$ and $P$ be, as above, the Hamiltonian and momentum of the theory,
and let $H_\pm =(H\pm P)/2$.
It is possible to find a linear combination $Q$ of the supercharges
$Q_\pm$ and $\b Q_\pm$ such that $Q$ is self-adjoint and $Q^2=H$.
Let $(-1)^F$ be the operator that counts the number of fermions
modulo two, and so anti-commutes with $Q$.  The object $\Tr (-1)^F
e^{-\beta H}$ is independent of $\beta$ by standard arguments
(involving a pairing of states at non-zero $H$ that comes from multiplication
by $Q$).  This quantity is the index of $Q$ (or more properly of the
piece of $Q$ that maps states even under $(-1)^F$ to odd states).
It is usually written simply
$$ \Tr (-1)^F. \eqn\juko$$
This index is independent of $r$ and $\theta$ (as long as we keep away from
the singularity at $r=\theta=0$) because of the discrete spectrum
for $H$ below a critical value; so it has the same value for
Landau-Ginzburg as for Calabi-Yau.

\REF\warner{A. Schellekens and N. Warner, ``Anomaly Cancellation And
Self-Dual Lattices, Phys. Lett. {\bf 181B} (1986) 339; K. Pilch, A.
Schellekens and N. Warner, ``Anomalies, Characters, and Strings,''
Nucl. Phys. {\bf B287} (1987) 317.}
\REF\wittenon{E. Witten, ``Elliptic Genera And Quantum Field
Theory,'' Commun. Math. Phys. {\bf 109} (1987) 525,
`'The Index Of The Dirac Operator In Loop
Space,'' in P. Landwebber, cited below. }
\REF\ellgen{P. Landwebber, ed., {\it Elliptic Curves And Modular Forms
In Algebraic Topology} (Springer-Verlag, 1988).}
A similar but much more refined invariant
can be constructed using the
left and right-moving $R$ symmetry of the theory.  Let $W_L=\oint J_L$
and $W_R=\oint J_R$ be the $R$ charges; and let
$$\eqalign{(-1)^{F_R} & = \exp(i\pi W_R) \cr
           (-1)^{F_L} & = \exp(i\pi W_L) .\cr } \eqn\pony$$
Then $(-1)^F=(-1)^{F_L}(-1)^{F_R}$.
Let $Q_R=(Q_++\b Q_+)/2$, so $Q_R$ anticommutes with $(-1)^{F_R}$
and commutes with $(-1)^{F_L}$, and $Q_R{}^2=H_+$.  By a standard
argument (a pairing of states at $H_+\not = 0$ that comes from multiplication
by $Q_R$), the quantity
$$\Tr (-1)^{F_R}q^{H_-}e^{i\theta W_L}
\exp(-\beta H_+)\eqn\jjdjjdd$$
is
independent of $\beta$; we denote this as
$$ F(q,\theta)= \Tr (-1)^{F_R}q^{H_-}e^{i\theta W_L}  . \eqn\omply$$
This quantity, which can be interpreted as the index of $Q_R$,
is essentially the elliptic genus of the model [\warner--\ellgen];
it has interesting modular properties that can be established from
its path integral representation.
Given the discrete spectrum for $H_+$ below a critical value,
$F(q,\theta)$ is invariant under  variation
of supersymmetric
parameters, so it is a topological invariant and moreover has the
same value for Calabi-Yau and Landau-Ginzburg.
(In the Landau-Ginzburg case, this function has not been previously
studied; it has quite interesting properties that will be explored elsewhere.)

In case $\sum_iQ_i$ is even but not zero,
$F(q,\theta)$ is still an invariant provided we restrict
$\theta$ to integer multiples of $2\pi\left(\sum_iQ_i\right)^{-1}$
(corresponding
to anomaly-free discrete $R$ symmetries) and provided we take due
account of the vacuum states at large $\sigma$.
The even-ness of $\sum_iQ_i$ is needed to ensure that $(-1)^{F_R}$ is not
anomalous; at low energies, it results in the
target space of the $r>>0$ sigma model being a spin manifold, a natural
requirement in defining  the elliptic genus.

\subsection{Continuation To Euclidean World-Sheet; Twisting}

So far in this paper, the two dimensional world-sheet has had
a Lorentzian signature.  For our remaining applications, a Euclidean
world-sheet is more natural.
We make the analytic continuation to a Euclidean signature world-sheet
by setting $y^0=-iy^2$; the line element $\d s^2=-(\d y^0)^2+(\d y^1)^2$
becomes $\d s^2=(\d y^1)^2+(\d y^2)^2$.

\REF\twist{E. Witten, ``Topological Sigma Models,'' Commun. Math. Phys.
{\bf 118} (1988) 411.}
\REF\eguchi{T. Eguchi and S.-K. Yang, ``$N=2$ Superconformal Models
As Topological Field Theories,'' Mod. Phys. Lett. {\bf A4} (1990)
1653.}
These linear sigma models not only are not topologically invariant;
they are not even conformally invariant, except (putatively) in the
infrared limit.  However, as for any $N=2$ supersymmetric models in two
dimensions, topological field theories can be constructed
by a simple twisting procedure [\twist,\witmir,\eguchi].
This involves substituting the stress tensor by
$$T_{\alpha\beta}\to T_{\alpha\beta}+{1\over 4}\left(
\epsilon_\alpha{}^\gamma\partial_\gamma (J_{R\beta}\pm J_{L\beta})
+\alpha\leftrightarrow\beta\right). \eqn\twisto$$
One obtains two models, depending on the choice of sign.  For
the $+$ and $-$ sign respectively one gets the so-called $B$ model
and $A$ model.  The $B$ model is anomalous unless $\sum_iQ_i=0$,
so our discussion of it is limited to that case.  There is no such
restriction for the $A$ model (but the $A$ model
has very special properties when $\sum_iQ_i=0 $, for only then the
instanton sums are infinite).

The point of the twisting procedure is that certain supercharges come to
have Lorentz spin zero if Lorentz transformations are defined using
the modified stress tensor.  These are $\b Q_+$ and $\b Q_-$ for
the $B$ model, and $Q_-$ and $\b Q_+$ for the $A$ model.
As these quantities are Lorentz invariant, they remain symmetries
when the models are formulated on an arbitrary Riemann surface $\Sigma$
of genus $g$.
Thus either the $A$ or the $B$ model, formulated on any surface $\Sigma$,
has a $(0|2)$ dimensional supergroup $F$ of symmetries.

Let $Q=\b Q_++\b Q_-$ or $Q= Q_- +\b Q_+$ for the $B$ or $A$ model;
then in either case $Q^2=0$.  It is natural to try to think of $Q$
as a sort of  BRST operator, considering physical states to
be the cohomology classes of $Q$ and the observables to be correlation
functions of $Q$ invariant vertex operators.
One can show, as in classical Hodge theory, that BRST cohomology classes
have $F$-invariant representatives;
this leads to some useful simplifications.

In either of the twisted models, one has $T_{\alpha\beta}=
\{Q,\Lambda_{\alpha\beta}\} $ for some $\Lambda_{\alpha\beta}$; this condition
ensures that the correlation functions of $Q$-invariant operators are
independent of the metric.  So the twisted models become
topological field theories; we call their observables the topological
correlation functions.
As explained in detail in
[\witmir], the topological observables of the $A$ and $B$ models
include as special cases
the $ac$ and $cc$ chiral rings of the untwisted model, which determine
the low energy Yukawa couplings of the string
compactification defined by the untwisted model.

Any perturbation of the Lagrangian
of the form $\int \d^2 y \,\d^4\theta\dots $
can be written as $\{Q,\dots\}$, so the topological observables are
invariant under such perturbations.  In particular, the topological
observables are independent of the gauge coupling $e$.
For other couplings the basic rule is as follows.  In the $B$ model,
any term in the Lagrangian that can be written as $\int \d\b\theta^+\dots$
or $\int\d\b\theta^-\dots$ is $\{Q,\dots\}$, while in the $A$ model
this is true for any term that can be written as $\int \d \theta^-\dots$ or
$\int \d\b\theta^+\dots$.

Looking at the definition of the superpotential
couplings,
$$L_W = \int \d^2 y \d\theta^+\d\theta^-\;\left.
W(\Phi_i)\right|_{\b\theta{}^+
=\b\theta^-=0} +\int\d^2y\d\b\theta{}^-\d\b\theta{}^+\;\left.
\b W(\b\Phi)\right|_{\theta^+
=\theta^-=0}, \eqn\pombo$$
we see that the observables of the $A$ model are independent of the
superpotential while the observables of the $B$ model vary holomorphically
with $W$.

Similarly, recalling the definition of the Fayet-Iliopoulos and theta
couplings,
$$L_{D,\theta}={it\over 2\sqrt 2}\int \d^2 y\d\theta^+\d\b\theta^-
\left.\Sigma\right|_{\theta^-=\b\theta{}^+=0}
-{i\b t\over 2\sqrt 2}\int\d^2y\d\theta^-\d\b\theta{}^+\left.\b\Sigma
\right|_{\theta^+=\b\theta{}^-=0},
\eqn\ombo$$
we see that the topological observables of the $B$ model are independent
of $r$ and $\theta$ while those of the $A$ model vary holomorphically in
$t$.

The $A$ model has a greatly enriched variant in which one
ignores $Q_-$ and considers $ \b Q_+$ as a BRST operator; thus physical
states or vertex operators are cohomology classes of $\b Q_+$.
(The space of physical states of the half-twisted model is infinite
dimensional; the elliptic genus \omply\ is essentially
the index of the $\b Q_+$ operator.)
This variant of the $A$ model is called the half-twisted model.
The half-twisted model is not a topological field theory; if this model
is formulated on a Riemann surface $\Sigma$, its correlation functions
are conformally invariant (even if the untwisted model was not) and
vary holomorphically with the complex structure of $\Sigma$.
The correlation functions of the half-twisted model are invariant under
change in $e$ and vary holomorphically with both $W$ and $t$, by
essentially the above arguments.

Since the interpolation between Calabi-Yau and Landau-Ginzburg
is obtained by varying $t$, we can draw some conclusions: the
topological observables of the $B$ model are the same for Calabi-Yau
and Landau-Ginzburg; for the $A$ and half-twisted models, Landau-Ginzburg
is an analytic continuation of Calabi-Yau.

\section{The Twisted Models In Detail}

We will now examine more closely the
$B$ and $A$ models.  (The half-twisted model can be considered
together with the $A$ model in what follows.)
The basic tool is a sort of fixed point theorem discussed in
[\witmir,\S5].
We recall that either
the $A$ or $B$ model has a $(0|2)$ dimensional supergroup $F$ of symmetries,
with fermionic generators that we will call $Q_\sigma$.
$F$-invariant representatives can be picked for all
of the vertex operators, so $F$ can be regarded as a symmetry of the
path integral.

The evaluation of the path integral for topological
correlation functions can be localized on the space $Z$ of
fixed points of $F$, that is the space of
points in field space for which $\{Q_\sigma,
\Lambda\}=0$
for every field $\Lambda$ and each $\sigma$.
If the $Q_\sigma $ have a simple zero along $Z$, then the path integral over
modes normal to $Z$ gives a simple factor of $\pm 1$ due to cancellations
between bosons and fermions.
Otherwise, the path integral reduces to an integral over a finite dimensional
space of directions in field space in which the $Q_\sigma$ vanish
to higher than first order.

\subsection{The $B$ Model}

In the $B$ model, by considering $\{\b Q_\pm,\b\lambda_\pm\}=0$,
one finds $D=v_{12}=0$.\foot{Recall the Wick rotation $y^0\to -iy^2$,
so $v_{01}\to -iv_{12}$.}  Let $\Phi_\alpha$ be the chiral
superfields with components $\phi_\alpha,\psi_\alpha,F_\alpha$ and charges
$Q_\alpha$.
By setting $\{\b Q_\pm ,\b\psi_{\alpha\pm}\}=0$, one gets $F=0$.
{}From $\{\b Q_\pm,\psi_{\alpha\pm}\}=0$, one learns that $D \phi_\alpha/D y^i
=0$, and that $Q_\alpha \phi_\alpha \sigma=0$ for all $\alpha$.  For
either $r>>0$ or $r<<0$, the equations $D=F_\alpha=0$ require
that the quantitites $Q_\alpha \phi_\alpha$ are not all zero, so the
last equation implies that $\sigma=0$.

\REF\vafatop{C. Vafa, ``Topological Landau-Ginzburg Models,''
Mod. Phys. Lett. {\bf A6} (1991) 337.}
In sum then, for the $B$ model, an $F$ fixed point is simply a constant
map from the world-sheet $\Sigma$ to the space of classical vacua.
For $r>>0$, the space of classical vacua is the target space $X$ of
the low energy sigma model.  In this case, the generators $Q_\sigma$
have simple zeros along their space $Z\cong X$ of zeroes, so the
path integral reduces to an integral over $X$.  This integral can
be analyzed as in [\witmir,\S4]; correlation functions can be evaluated
in terms of periods of differential forms on $X$.
For $r<<0$, the space of classical vacua is a point, so the $Q_\sigma$
have only one zero, but this zero is degenerate;
the path integral reduces to a finite dimensional integral analyzed
by Vafa [\vafatop].

\subsection{The $A$ Model}

For the $A$ model, the equations $\{Q_\sigma,\lambda_\pm\}=0$
and $\{Q_\sigma,\b\lambda_\pm\}=0$ give
$$\eqalign{  D+v_{12} & = 0 \cr
             \d \sigma  & = 0 .\cr}\eqn\nurgog$$
The equations $\{Q_\sigma,\psi_{\alpha\pm}\}=0$ and $\{Q_\sigma,
\b\psi_{\alpha\pm}\}=0$ similarly give
$$\eqalign{  \left({D\over Dy^1}+i{D\over D y^2}\right)\phi_\alpha & = 0 \cr
        F_\alpha & = 0 \cr
        Q_\alpha \phi_\alpha\sigma & = 0. \cr}  \eqn\turgog$$
The last equation in \turgog\ and the last equation in \nurgog\
imply, together, that $\sigma=0$, since for the models of interest
the other equations do not permit $Q_\alpha \phi_\alpha$
to vanish identically for each $\alpha$.
Altogether, the surviving equations are
$$\eqalign{  \left({D\over Dy^1}+i{D\over D y^2}\right)\phi_\alpha & = 0 \cr
             F_\alpha & = 0 \cr
             D+v_{12} & = 0. \cr} \eqn\rovero$$

{}From the identities
$$\eqalign{ \int \d^2y D_i\b \phi_\alpha D^i \phi_\alpha= & \int
\d^2y(D_1-iD_2)
\b \phi_\alpha
 (D_1+iD_2)\phi_\alpha -\int \d^2y Q_\alpha
\b \phi_\alpha \phi_\alpha v_{12} \cr
}\eqn\bufuf$$
$$\eqalign{  {1\over 2e^2}\int \d^2 y(v_{12}{}^2+D^2)
 = & {1\over 2e^2}\int \d^2 y(v_{12}+D)^2
-{1\over e^2}\int \d^2 y \,\,\,D v_{12} \cr} \eqn\nufuf$$
and $D=-e^2(\sum_\alpha Q_\alpha\b \phi_\alpha \phi_\alpha -r)$, we find
that if $D+v_{12}=(D_1+iD_2)\phi_\alpha=0$, then
$$\int\d^2y\left( \sum_\alpha D_i\b \phi_\alpha D^i\phi_\alpha
+{1\over 2e^2}\left(v_{12}{}^2+D^2\right)\right) =-r\int\d^2y \,\,v_{12}.
\eqn\wimomo$$
The left hand side coincides with the bosonic part of the Lagrangian
of the theory, modulo $\sigma=F_\alpha=0$, and apart from the theta term.
So the action of a solution of \rovero\
$$L =-2\pi i t N  \eqn\uvvu$$
where $N$ is the instanton number
$$N=-{1\over 2\pi }\int\d^2y\,\, v_{12}, \eqn\jiko$$
and $t=ir+\theta/2\pi$; the substitution of $r$ by $-it$
takes account of the theta term
in the action.

Since the left hand side of
\wimomo\ is positive definite, it follows that solutions
of \rovero\ necessarily have $N\geq 0$ if $r>0$, and $N\leq 0$ if $r<0$.
The ``instanton expansion'' for the evaluation of any topological
observable is therefore an expansion in instantons,
of the general form
$$\sum_{k\geq 0} a_k\exp(2\pi i k t) ,\eqn\uccip$$
for $r\to \infty$, or an expansion in anti-instantons, of the general form
$$\sum_{k\geq 0}b_k\exp(-2\pi i k t) ,\eqn\nuccip$$
for $r\to -\infty$.

A standard vanishing theorem (a line bundle of negative degree
cannot have a non-zero holomorphic section) says that
if $(D_1+iD_2)\phi_\alpha=0$, then
$$ \phi_\alpha=0 ~~~{\rm unless}~~
         {\rm sign}(Q_\alpha) ={\rm sign}(N). \eqn\mico$$
This follows from \bufuf\ if one uses the invariance of
$(D_1+iD_2)\phi=0$ under complex gauge transformations
(discussed below) to set $v_{12}={\rm constant}$.

Let us now look more closely at the structure of the instantons and
anti-instantons, for the case relevant to the simplest form of the
C-Y/L-G correspondence: the case in which there are $n$ chiral superfields
$S_i$ of charge 1, and one chiral superfield $P$ of charge $-n$,
and the superpotential is $W=P G(S_i)$.

\subsection{$r<<0$}

The case of $r<<0$ is easier, so we consider it first.
In view of \mico, $s_i=0$, but $p\not=0$.  Vanishing of $s_i$ ensures
that $F_\alpha=0$.
The remaining equations are
$$\eqalign{ (D_1+iD_2) p & =  0 \cr
             v_{12}      & = e^2(-n|p|^2-r).\cr}     \eqn\judop$$
These are the equations of  the Nielsen-Olesen abelian vortex line,
and the qualitative properties of the solutions are well known.

The $p$ field is massive,
with a mass proportional to $e\sqrt r$, and Compton wavelength
$\lambda\sim 1/e\sqrt r$.
Because the $p$ field has charge $-n$, the solutions of \judop\
can have instanton
number $-k/n$ for arbitrary integer $k\geq 0$.
In an anti-instanton field of instanton number $-k/n$,
the $p$ field vanishes at
$k$ (generically distinct)
points $x_1\dots x_k \in \Sigma$, and the energy density of
the anti-instanton field is concentrated within a distance of order
$\lambda$ from those points.  To a low
energy observer, probing distances much longer than $\lambda$,
the field looks like it is concentrated at $k$ points; it looks like
a superposition of $k$ point anti-instantons each of instanton number $-1/n$.

Around one of the $x_a$, where the $p$ field has a simple zero,
it changes in phase by $2\pi$.  The $s_i$, of charge $1$, therefore
change in phase by $-2\pi /n$.  Therefore, from the point of view of
the low energy effective theory of the massless fields $s_i$,
it looks like ``twist fields,'' about which the $s_i$ change in phase
by $-2\pi/n$, have been inserted at the $k$ points $x_a$.  Of course,
in a global situation, on a compact Riemann surface $\Sigma$, the total
change in phase of the $s_i$ must be a multiple of $2\pi$, so $k$ must
be divisible by $n$.
This was to be expected; it is the standard quantization of the instanton
number in the presence of fields of charge $1$.

We see therefore the interpretation of the parameters $r,\theta$ in the
Landau-Ginzburg theory of $r<<0$.  There is a gas of twist fields,
with a chemical potential such that an amplitude with $k$ twist fields
receives a factor of
$$  \exp(-2\pi i tk/n).\eqn\ududucc$$

For vacuum amplitudes, or amplitudes with insertions of ordinary
(untwisted) fields, the number $k$ of twist fields in the gas must
be divisible by $n$.  It is however known in the theory of the Landau-Ginzburg
model that the $A$ model has BRST invariant observables
in the twisted sectors.  A correlation function with, say, $s$ twist
fields receives contributions only from values of $k$ such that $k+s$ is
divisible by $n$.  For given $s$, let $k_0$ be the smallest non-negative
integer with $k_0+s$ divisible by $n$.  A correlation function
$\langle T_1\dots T_s\rangle$ of $s$ twist fields is then proportional
for $r\to -\infty$ to
$$\exp(-2\pi i t k_0/n) \eqn\jugug$$
and in particular vanishes at $r=-\infty$ unless $k_0=0$.
This means that at $r=-\infty$ there is a selection rule not present
otherwise: $\langle T_1\dots T_s\rangle=0$ unless $s$ is divisible by $n$.
This selection rule corresponds to a quantum $\Z_n$ symmetry that holds
only at $r=-\infty$.

To recapitulate, let us emphasize that for $r<<0$, the low energy
theory is a Landau-Ginzburg theory, which does not have (smooth) instantons.
The construction of smooth instanton solutions depends on the massive
fields; the instanton solutions are smooth at a fundamental level,
but look like point singularities in the low energy theory.

\subsection{$r>>0$}

If conversely  $r>>0$,
we are dealing with instantons rather than anti-instantons;
\mico\ forces $p=0$.  With the superpotential being $W=PG(S_i)$,
the equations to be obeyed are
$$\eqalign{ (D_1+iD_2)s_i & = 0 \cr
                 G(s_i)   & = 0 \cr
                  v_{12}  & = e^2(\sum_i|s_i|^2-r) .\cr} \eqn\huccox$$

We will presently analyze the import of these equations, but first,
without any mathematical formalism, let me state the result that will
arise.  The low energy theory for $r > >0$ is a sigma model with target
space a Calabi-Yau hypersurface $X\subset \CP^{n-1}$.  We should certainly
expect to find the instantons of the sigma model, that is the holomorphic
maps of $\Sigma$ to $X$.  However (just as for $r<<0$), we will also
find additional instantons that are perfectly smooth objects at the fundamental
level, with the massive fields present, but look like singular objects
in the effective low energy theory.   These additional instantons
have structure varying on a scale proportional
to the Compton wavelengths of the massive fields (and other structure
varying more smoothly).

The presence of these additional ``singular'' instantons means that the
instanton expansion of the linear sigma
model studied in this paper cannot be expected
to coincide with the standard instanton expansion of the conformally
invariant nonlinear sigma model to which it reduces at low energies.
However, conventional renormalization theory tells us how they
will be related.  We should think about the effects
of integrating  out the point-like instanton structures
(along with all other effects of the massive fields) to get an effective
theory of the massless fields.  This theory must be a sigma model
of some kind, with some effective values of $r$ and $\theta$ and some effective
defining equation $G=0$.  (All other parameters are believed to be
``irrelevant'' in the sense of the renormalization group.)  Since the
$A$ model observables do not depend on the particular polynomial $G$,
the effect of integrating out the point instantons should be simply
to change the effective values of $r$ and $\theta$.
Thus, the instanton expansion of the linear sigma model
should differ from the conventional
instanton expansion of the nonlinear
sigma model by a redefinition of the variable $t$.

\REF\vortexmod{S. Bradlow, ``Special Metrics And Stability For Holomorphic
Bundles With Global Sections,'' J. Diff. Geom. {\bf 33} (1991) 169;
S. Bradlow and G. Daskapoulos, ``Moduli Of Stable Pairs For Holomorphic
Bundles Over Riemann Surfaces,'' Int. J. Math. {\bf 2} (1991) 477; O.
Garcia-Prada, ``Dimensional Reduction Of Stable Bundles, Vortices,
And Stable Pairs,'' in preparation.}
Now I come to the mathematical analysis of \huccox. Obviously,
these equations are invariant under gauge transformations
$$ s_i\to e^{i\lambda} s_i,~~~v\to v-\d \lambda, \eqn\bilbop$$
with real $\lambda$.  Actually, it is easy to see that the first two
equations in \huccox\ are invariant under \bilbop\ even if $\lambda$
is {\it complex}.  It is a very beautiful fact [\vortexmod]
that the last equation in \huccox\ can be interpreted as a condition
that fixes the complex gauge invariance of the first two equations.
In other words, the space of solutions of the first two equations,
up to a complex gauge transformation, is the same as the space of solutions
of the set of three equations, up to a real gauge transformation.  (This
can be regarded as an infinite dimensional analog of the relation
between complex and symplectic quotients summarized in the next
section.)

So we can study simply the first two equations in \huccox, up to
complex gauge transformations.  To make the analysis particularly
simple, we will assume also that the world-sheet $\Sigma$ is of genus
zero (the main case in standard applications of the $A$ model).
This leads to simplification because any complex line bundle on $\Sigma$
is ${\cal O}(k)$ for some $k$.  If $u,v$ are homogeneous coordinates
for $\Sigma\cong \CP^1$, then a holomorphic section of ${\cal O}(k)$
is just a holomorphic function of $u,v$ that is homogeneous of degree $k$.
I will later use the same name ${\cal O}(k)$ to denote the analogous
line bundle over $\CP^{n-1}$.

Up to a complex gauge transformation, the first equation in \huccox\
simply says that the $s_i$ are holomorphic sections of ${\cal O}(k)$,
$k$ being the instanton number.  Thus $s_i=s_i(u,v)$ are polynomials
in $u$ and $v$, homogeneous of degree $k$, explicitly
$$s_i(u,v)=s_{i,k}u^k+s_{i,k-1}u^{k-1}v+\dots + s_{i,0}v^k.\eqn\mcmcmmm$$
The overall scaling
$$ s_i(u,v) \to t s_i(u,v),~~~ t\in \C^* \eqn\djd$$
corresponds to the complex gauge transformations
with constant gauge parameter; these are isomorphisms of ${\cal O}(k)$.
The second equation in \huccox\
says that
$$G(s_1(u,v),\dots, s_n(u,v))=0.\eqn\nurgo$$
The moduli space of solutions of \huccox\ is the space of degree $k$
polynomials $s_1(u,v),\dots, s_n(u,v)$ obeying \nurgo, modulo \djd.

Now let us consider the moduli space of the sigma model, that is the
moduli space of degree $k$ holomorphic maps
$\Phi:\Sigma\to X$, $X$ being the hypersurface $G=0$ in
$\CP^{n-1}$.
This sigma model moduli space is easy to describe.
$\Phi$ having degree $k$ means by definition
that $\Phi^*({\cal O}(1))={\cal O}(k)$.
The homogeneous coordinates $s_i$ of $\CP^{n-1}$ are holomorphic
sections of ${\cal O}(1)$, so $\Phi^*(s_i)$ are holomorphic sections
of ${\cal O}(k)$.  This means that any degree $k$ holomorphic map
$\Phi:\CP^1\to \CP^{n-1}$ is of the form
$$(u,v)\to \left(s_1(u,v),\dots,s_n(u,v)\right) \eqn\hdond$$
with degree $k$ polynomials $s_1(u,v),\dots s_n(u,v)$ which must obey
\nurgo.  Since in $\CP^{n-1}$
the $s_i$ are not permitted to vanish simultaneously, these polynomials
must have no common zeroes.  Conversely, if the $s_i$
obey \nurgo\ and
have no common zeroes, the formula \hdond\ describes a degree $k$
holomorphic map $\CP^1\to \CP^{n-1}$.

The relation between the two moduli spaces is now clear.  Every sigma
model instanton comes from a solution of \huccox, but \huccox\ has
extra solutions, in which the $s_i(u,v)$ do have common zeroes for
some $u,v$, not both zero.  These extra solutions are
smooth objects in the underlying gauge theory, but near the common
zeroes of the $s_i$ they have a scale of variation
of order $1/e\sqrt r$; this is the ``point-like'' behavior promised
in our introductory remarks.
These instantons show ``point-like'' behavior only near the common
zeroes of the $s_i$;  elsewhere they look like sigma
model instantons.

One might think that the condition that the $s_i$ have common zeroes
would be obeyed only very exceptionally, so that the point-like instantons
would perhaps be irrelevant.  As a preliminary to seeing that this is
not so, let us work out the dimension of the instanton moduli space.
Equation \nurgo\ asserts
the vanishing of a polynomial in $u,v$ homogeneous of degree $kn$,
which therefore has an expansion $G(u,v)=a_{kn}u^{kn}+a_{kn-1}u^{kn-1}v
+\dots +a_0v^{kn}$.  The $kn+1$ $a$'s can be written out explicitly
as polynomials of degree $n$ in the $(k+1)n$ coefficients $s_{i,r}$.
Allowing also for the scaling relation \djd,
the dimension of the instanton moduli space is  expected to be
$(k+1)n-(kn+1)-1
=n-2$ if everything is generic.  It is known that this is the actual
dimension of the sigma model moduli space, for generic $G$.  (In the
usual case of target space a quintic three-fold in $\CP^4$, $n-2=3$,
and the dimension is reduced to zero after dividing by the action of
$SL(2,\C)$ on $\Sigma$.)

If now $s_i(u,v)$ are a collection of degree $k$ polynomials, obeying
\nurgo\ and having no common zeros, then $\widetilde s_i(u,v)
=(\alpha u+\beta v)s_i(u,v)$ are a collection of degree $k+1$ polynomials,
obeying \nurgo, and with a common zero where $\alpha u+\beta v=0$.
After allowing for an irrelevant scaling of $\alpha,\beta$,
the $\widetilde s_i$ depend on one more parameter than
the $s_i$; thus, while the sigma model instantons depend on $n-2$
parameters, the point-like instantons with one common zero depend
on $n-1$ parameters.
(And the point-like instantons with $w$ common zeros depend on $n-2+w$
parameters.)
The point-like instantons are in no way
exceptional, and they must be considered in integrating out the massive
fields to get an effective action for the massless fields.
Conventional renormalization lore suggests, as noted earlier, that their
effect is to induce a reparametrization of the variable $t$.

\chapter{SOME GEOMETRICAL BACKGROUND}

\REF\gks{V. Guillemin and S. Sternberg, ``Geometric Quantization
And Multiplicities Of Group Representations,'' Invent. Math. {\bf 67}
(1982) 515.}
\REF\mumford{D. Mumford and J. Fogarty, {\it Geometric Invariant Theory},
(Springer, 1982).}
The construction of \S3 has
many generalizations, some of which we will explore in \S5.
First, however, I want to explain some geometrical background that
sheds some light on the subject.  These facts are not
strictly necessary for reading \S5, but clarify some of the issues,
especially when we come to considering changes of topology in \S5.5.
In effect we will be explaining, in an elementary and {\it ad hoc}
way, the relation (see [\gks] or [\mumford, p. 158])
between symplectic and holomorphic quotients.

Consider the manifold $Y=\C^{n+1}$ with coordinates $s_1,\dots , s_n$
and $p$, and with the $\C^*$ action
$$\eqalign{s_i & \to \lambda s_i     \cr
           p & \to \lambda^{-n}p     \cr} \eqn\jojoj$$
for $\lambda\in \C^*$.
We want to form a quotient of $Y$ by $\C^*$.
The relevance of this will gradually become clear.
Since $\C^*$ acts freely on $Y$ minus the origin $O$, one might
think that one could straightforwardly form a reasonable quotient
$(Y-O)/\C^*$.  For  a free action of a compact group, a reasonable
quotient always exists, but the story is quite different for non-compact
groups.

Let $P$ be a point in $Y$ with $p=0$, and let $P'$ be a point in $Y$ with
$s_1=\dots =s_n=0$.  Under the $\C^*$ action, $P$ can be brought arbitrarily
close to the origin (by taking $\lambda\to 0$), and the same can be done
to $P'$ (by taking $\lambda\to\infty$).  Even if we delete the origin
in forming the quotient, any neighborhood of $P$ has a $\C^*$ orbit that
intersects any neighborhood of $P'$.  This means that
the set theoretic quotient $(Y-O)/\C^*$, with its natural induced topology,
is not a Hausdorff space.

Let $Y_1$ be the subset of $Y$ with $p=0$ and the $s_i$ not all 0;
let $Y_2$ be the subset $s_i=0$, $p\not=0$.
The $\C^*$ orbits that come arbitrarily close to the origin
are (apart from $O$ itself) the orbits in $Y_1$ or $Y_2$.
If $\widetilde Y
=Y-(Y_1\cup Y_2 \cup O)$, then since $\widetilde Y$ only contains ``good''
$\C^*$ orbits (that are closed and bounded away from the origin),
and $\C^*$ acts freely on $\widetilde Y$,
the quotient space $\widetilde Y/\C^*$ is a manifold,
just as if $\C^*$ were compact.
Deleting $Y_1,$ $Y_2$, and $O$ is too crude, however; by suitably including
some of the ill-behaved orbits one can obtain natural partial
compactifications of $\widetilde Y/\C^*$.

\REF\newstead{P. Newstead,  {\it Introduction To Moduli Problems
And Orbit Spaces} (Tata Institute, 1978).}
In the simple case that we have considered, it is not hard to guess how
to do this.  And there is a systematic theory (geometric invariant theory
[\mumford];
for an introduction see Newstead's book [\newstead]).  Instead of
studying this situation purely mathematically, let us return to the
physics problem of \S3.1.
In that problem, $Y$ was endowed with the Kahler metric $\d \tau^2
=\sum_i|\d s_i|^2+|\d p|^2$.  This Kahler metric is not $\C^*$
invariant, so $\C^*$ is not a symmetry group of the $N=2$ supersymmetric
field theory
studied in \S3.  The metric is, however, invariant under the maximal
compact subgroup $U(1)\subset \C^*$ (whose action is given
by \jojoj\ with $|\lambda|=1$).  This $U(1)$ was used in \S3 as the gauge
group.
The Kahler form associated with
the Kahler metric $\d\tau^2$ is $\omega=-i\sum_i\d\bar s_i\wedge \d s_i
-i\d\bar p\wedge \d p$.
The $U(1)$ action on $Y$ preserves this Kahler form.  Since $Y$ is
simply connected, there inevitably is a Hamiltonian function that generates
by Poisson brackets the $U(1)$ action on $Y$; it is simply
$$\widetilde D=-{D\over e^2}=\sum_i |s_i|^2-n|p|^2-r, \eqn\nurgo$$
with $r$ being an arbitrary constant.
The $D$ function that played such an important role in \S3 is thus
simply the generator, in this sense, of the gauge group.

\REF\dh{J. J. Duistermaat and G. J. Heckman, ``On The Variation
In The Cohomology In The Symplectic Form Of The Reduced Phase
Space,'' Invent. Math. {\bf 69} (1982) 259.}
In \S3, it was natural to set $D=0$, so as to minimize the energy, and
then divide by the gauge group.  The combined operation of setting $D=0$
and then dividing by $U(1)$ gives what is called the symplectic quotient
of $Y$ by $U(1)$.  The symplectic quotient of $Y$ by $U(1)$, often
denoted $Y\catquot
U(1)$, depends on $r$, of course.  Even the topology of $Y\catquot U(1)$
changes when $r$ passes through certain distinguished values at which $Y
\catquot U(1)$
is singular -- in our example
this occurs only at $r=0$.
And when the topology is not changing, the sympectic structure of $Y\catquot
U(1)$
still depends on $r$.
$Y\catquot U(1)$ has a natural symplectic structure
obtained by restricting $\omega$ to
$D=0$ and then projecting to the quotient $Y\catquot U(1)=
\{D=0\}/U(1)$.  This natural symplectic structure
determines an element of $H^2(Y\catquot U(1),\R)$ which
depends linearly on $r$; this is related to the Duistermaat-Heckman
integration formula [\dh].

However, if we consider not $Y$ but $\widetilde Y$,
certain simplifications arise: $\widetilde
Y\catquot U(1)$ is in a certain sense naturally
independent of $r$.  To see this, we consider the $D$ function restricted
to a particular $\C^*$ orbit.  It is
$$\widetilde D(\lambda) =|\lambda|^2\sum_i|s_i|^2-n|\lambda|^{-2n}|p|^2-r.
\eqn\jcuc$$
Because we are in $\widetilde Y$, the coefficients of $|\lambda|^2$ and
$|\lambda|^{-2n}$ are both non-zero.  $\widetilde D(\lambda)$ is a monotonic
function of $|\lambda|$ which goes to $+\infty$ for $\lambda\to \infty$
and to $-\infty$ for $\lambda\to 0$.  It follows that $\widetilde D(\lambda)=0$
for a unique value of $|\lambda|$.

The given $\C^*$ orbit contributes, of course, precisely one point to
the quotient $\widetilde Y/\C^*$.  It also contributes precisely one point to
$\widetilde Y\catquot
U(1)$, since $|\lambda|$ is uniquely determined by requiring
$D=0$, and the argument of $\lambda$ is absorbed in the $U(1)$ action.
Therefore, $\widetilde Y\catquot
U(1)$ coincides naturally with $\widetilde Y/\C^*$.

{}From the symplectic point of view, there is no mystery about how
to include the bad points -- the origin $O$ and the points in $Y_1$ and
$Y_2$.
The result, however, depends on $r$:

(i) For $r>0$, setting $D=0$ is possible in $Y_1$ but not for the other
bad points.  The symplectic quotient $Q=Y_1\catquot
U(1)$ is a copy of $\CP^{n-1}$,
with Kahler form proportional to $r$.  So $Y\catquot U(1)$ is the union
of $\widetilde Y/\C^*$ with this $\CP^{n-1}$.

(ii) For $r=0$, the origin is the only bad point with $D=0$.  Its
sympectic quotient is a single point, and $Y\catquot U(1)$ is the union of
$\widetilde Y/\C^*$ with this point.

(iii) For $r<0$, setting $D=0$ is possible in $Y_2$ but not for the other
bad points.  The symplectic quotient $Y_2\catquot U(1)$ is a single point,
and $Y\catquot U(1)$ is the union of $\widetilde Y/\C^*$ with this point.

Since $Q=Y_1\catquot U(1)$ is the same as $Y_1/\C^*$ (both being $\CP^{n-1}$),
and similarly $Y_2\catquot U(1)=Y_2/\C^*$ (both being a point),
we can restate these results holomorphically.  Let us do this in detail
for $r>0$ and
for $r<0$:

(i) For $r>0$, $Y\catquot U(1)$ is the same as $Z=Y'/\C^*$, where
$Y'=\widetilde Y\cup Y_1$
is the union of the points at which the $s_i$ are not all zero.
$Z$ is fibered over $\CP^{n-1}$ (by forgetting $p$).  The fiber is a copy
of the complex plane, parametrized by $p$.  $Z$ is therefore a complex
line bundle over $\CP^{n-1}$; in fact, in view of the transformation
law in \jojoj, $Z$ is the total space of the line bundle ${\cal O}(-n)$.
$Z$ is actually a non-compact Calabi-Yau manifold.  To see this,
begin with the equation $\sum_i Q_i=0$ that played such an important
role in \S3 (the $Q_i$ being here the exponents in \jojoj).  This
equation ensures that the $n+1$ form
$\Theta=\d s_1\wedge\dots\wedge \d s_n\wedge \d p$
is $\C^*$ invariant.  Contracting $\Theta$ with the vector field generating
the $\C^*$ action, we get an everywhere non-zero holomorphic $n$-form
$\Theta'$
whose restriction to $Y'$
is the pullback of a holomorphic $n$-form on $Z=Y'/\C^*$.

(iii) For $r<0$, $Y\catquot U(1)$ is the same as $Y''/\C^*$ where
$Y''=\widetilde Y\cup  Y_2$ is the region with $p\not=0$.  We can therefore
use the $\C^*$ action on $(s_i,p)$ to set $p=1$.  This leaves a residual
invariance under the subgroup of $U(1)$ defined by $\lambda^n=1$.
Dividing by this group, $Y\catquot
U(1)$ for $r<0$ is the same as $Z'=\C^n/\Z_n$.
This is again a non-compact Calabi-Yau manifold, by virtue of $\sum_iQ_i=0$.

There are two main conclusions:

(a) $Y\catquot U(1)$ carries a natural complex structure for any $r$.
It is evident ``physically'' that this must be so, since the low energy
limit of the models studied in \S3 is (if we set the superpotential to zero)
an $N=2$ sigma model with this target space.  (The analog of this reasoning
for $N=4$ leads to the celebrated hyper-Kahler quotient [\hitch].)

(b) The various complex manifolds $Y\catquot U(1)$ for $r$
positive, negative, or zero are all equivalent to each other on dense
open sets, since on a dense open set they all coincide with $Y'/\C^*$.
The technical term for this
is that these complex manifolds are birationally equivalent.  In fact,
we can be more specific.  $Z$ is obtained by blowing up the origin
in $Z'$ -- replacing the origin in $Z'=\C^n/\Z_n$ (which originated
as $Y_2\catquot U(1)$) with a copy of $\CP^{n-1}\cong Y_1\catquot U(1)$.

Though I have illustrated the ideas in a very special case, these
conclusions are general [\gks,\mumford].
Given an action of a  compact group $G$ on, for instance,
$Y= \C^n$ or on
a projective variety (endowed with a choice of Kahler metric)
the various possible symplectic quotients $Y\catquot
G$  (obtained from different
choices of $D$ functions) carry natural
complex structures and are all birationally equivalent to one another.
Indeed, they can all be identified on a dense open set with $\widetilde Y/
G_{\C}$, where $\widetilde Y$ is a suitable dense open set in $Y$, and
$G_\C$ is the complexification of $G$.

The technique of \S3 is really a technique
for interpolating between sigma models with birationally equivalent
target spaces, obtained by varying the $D$ functions.
This technique can be applied in the absence of any superpotential,
in which case it leads directly to relations between some sigma models
with birationally equivalent targets.  We will pick
up this theme in \S5.5 in connection with transitions in the topology
of space-time.

On the other hand, if one begins with a $\C^*$-invariant
holomorphic function $W$ on
$Y$ -- serving as a superpotential -- then $W$ will descend to a
holomorphic function  $\widehat W$ -- serving as a superpotential
-- on any of the symplectic quotients $Y\catquot
G$.  Thus, any relation between
sigma models with birationally equivalent targets obtained by the method
of \S3 extends to the case in which
these spaces are endowed with ``common'' superpotentials, that is
superpotentials that come from a common underlying gauge invariant
function on $Y$.
In general, birational equivalence is the condition under
which it makes sense to speak in this way
of the ``same'' holomorphic function on different
complex manifolds.

For instance, in the case treated in \S3,
the underlying superpotential was $W=pG(s_i)$, with $G$ being homogeneous
of degree $n$.  For $r>0$, $W$ restricts and descends to a holomorphic
function $\widehat  W$ on the symplectic quotient $Z$.
The effective theory after integrating out the massive gauge
multiplet is a theory with target space $Z$
and superpotential $\widehat W$; by essentially the computations of \S3,
it reduces at low energies to the sigma model of the Calabi-Yau hypersurface
$G=0$ in $\CP^{n-1}\subset Z$.

On the other hand, for $r<0$, $W$ restricts and descends to the holomorphic
function $G(s_i)$ on $Z'=\C^n/\Z_n$.  This is the superpotential of the
familiar Landau-Ginzburg orbifold.

In this way, the birational equivalence between $Z$ and $Z'$
lies behind         the C-Y/L-G correspondence.
The basic relation is the one between quantum field theory on $Z$ and on $Z'$;
the C-Y/L-G correspondence arises upon examining this relation
in the presence of a particular common superpotential.

\chapter{GENERALIZATIONS}

Our task in this section is to describe some generalizations
of the Calabi-Yau/Landau-Ginzburg correspondence.
We will first consider fairly immediate generalizations, involving
(i) hypersurfaces in a weighted
projective space; (ii) hypersurfaces in products of projective
spaces and in more general toric varieties; (iii)
hypersurfaces in Grassmannians; (iv) intersections of hypersurfaces
in any of those spaces.
In each case, the discussion will be rather brief because most
of the discussion in \S3 carries over with obvious modifications.
Finally, in \S5.5, we will discuss the occurrence in this framework
of transitions between manifolds of different topology.
The framework of \S4 is useful background, especially for \S5.5.

\section{Hypersurfaces In Weighted Projective Space}

\REF\schim{M. Kreuzer and H. Skarke, ``On The Classification
Of Quasi-Homogeneous Functions,'' ``No Mirror Symmetry In
Landau-Ginzburg Spectra'' (preprints, 1992).}
The most obvious slight generalization of what we have already done
in \S3 is the following.
Consider a $U(1)$ gauge theory with $n$ chiral superfields $S_i$,
$i=1\dots n$, of charge $q_i$, and one more chiral superfield
$P$ of charge
$-\sum_i        q_i$.   By rescaling the $U(1)$ generator, one can assume
that the $q_i$ are relatively prime.

The above choice of charges
ensures that $\sum_iQ_i=0$, a condition which as we saw
in \S3 leads to anomaly-free left- and right-moving $R$ symmetry
and hence to Calabi-Yau manifolds at low energies.
In our subsequent examples, we will always ensure in a similar way
that $\sum_iQ_i=0$, for each $U(1)$ charge, without pointing this out
explicitly in each case.

We take the superpotential to be
$$W = P G(S_i), \eqn\ompo$$
where $G$ is a polynomial of charge $q=\sum_iq_i$, transverse
in the sense that the equations
$${\partial G\over \partial S_1}=\dots={\partial G\over \partial S_n}
=0 \eqn\hompo$$
are obeyed only at $S_1=\dots=S_n=0$.  (Existence of such a $G$ is a
severe restriction on the $q_i$, analyzed in [\schim],
but if any $G$ of charge $q$ obeys this condition, the generic one does.)
The part of the Lagrangian containing the chiral superfields is thus
$$\int\d^2y\d^4\theta\left(\sum_i \b S_ie^{2V}S_i+\b Pe^{-2qV}P\right)
-\int\d^2y\d^2\theta PG(S_i) - \int\d^2y\d^2\b\theta \,\,\b P\,\,\b G(\b S_i).
\eqn\omco$$
The potential energy is much as before
$$U(A_i,\sigma)={1\over 2e^2}D^2+\sum_\alpha|F_\alpha|^2
+2\b\sigma \sigma\left(\sum_iq_i{}^2|s_i|^2
+q^2|p|^2\right),
\eqn\bumbop$$
with
$$\eqalign{ D & =-e^2\left( \sum_i q_i|s_i|^2-q|p|^2 -r\right) \cr}\eqn\tumbo$$
and
$$\sum_\alpha|F_\alpha|^2=|G|^2+|p|^2\sum_i\left|{\partial G\over \p s_i}
\right|^2. \eqn\irmob$$
Since we will have to write similar formulas many
times, we will adopt a few conventions.  The symbol $\sum_\alpha$
will always involve a sum over all chiral superfields $\Phi_\alpha$,
with auxiliary fields $F_\alpha$.
The terms in the potential involving $\sigma$ always have the effect
of giving $\sigma$ a mass and forcing its vacuum expectation value to
vanish (away from phase boundaries; the behavior near phase boundaries
is always as explained in \S3.2).  Knowing this, $\sigma$ can be dropped
from the discussion, and the potential can be written only at $\sigma=0$.

The analysis proceeds as in \S3.1.
One step in finding the ground state structure is to set $D=0$ and
divide by the gauge group
$U(1)$ (forming the symplectic quotient of $\C^{n+1}$ by
$U(1)$, in the language of \S4).
If $r>>0$, the locus $D=0$ divided
by $U(1)$ is a copy of the weighted projective space
${\bf WCP}^{n-1}_{q_1,\dots,q_n}$, with Kahler class proportional to $r$.
The low energy theory is a sigma model whose target space
is the hypersurface $G=0$ in that weighted projective
space.
This hypersurface is a Calabi-Yau manifold because of the underlying
$R$-invariance.
The assumption that the $q_i$ are relatively prime means
that (away from singularities of the weighted projective space)
the gauge symmetry is completely broken.

On the other hand, for $r<<0$, the field $p$ gets a vacuum expectation
value, breaking the gauge group down to $\Z_q$.
There is a unique vacuum, with $s_i=0$ and $p=\sqrt {-r/q}$, up to
a gauge transformation.  The low energy theory is a $\Z_q$ orbifold
with superpotential $G(s_1,\dots,s_n)$.  We have recovered
the standard correspondence between Calabi-Yau hypersurfaces in
weighted projective space and Landau-Ginzburg orbifolds.

\section{Hypersurfaces In Toric Varieties}

Now I want to consider a case in which we will discover something
new, instead of just recovering known results.  To avoid cluttering
up the notation, I begin by considering a very special case
of a hypersurface in a product of projective spaces, say $\CP^{n-1}
\times \CP^{m-1}$.

First of all, to describe $\CP^{n-1}\times \CP^{m-1}$, we consider
a $U(1)\times U(1)$ gauge theory with two vector superfields,
say $V_1$ and $V_2$, with gauge couplings $e_1$ and $e_2$.
There are therefore two independent Fayet-Iliopoulos terms, in components
$$L_{\mit D}=-\int\d^2y  \left(r_1D_1+r_2D_2\right).\eqn\konko$$
We introduce superfields $S_i,\,\,i=1\dots n$, of charges $(1,0)$,
$T_j,\,\,j=1\dots m$, of charges $(0,1)$, and one more superfield $P$
of charges $(-n,-m)$.
The kinetic energy of these fields is thus
$$\int\d^2y\d^4\theta\left(\sum_i\b S_ie^{2V_1}S_i
+\sum_j\b T_je^{2V_2}T_j+\b Pe^{-2nV_1-2mV_2}P\right).  \eqn\occo$$
For the superpotential, we take
$$W=P G(S_i,T_j), \eqn\tocco$$
where $G$ is a polynomial homogeneous of degree $n$ in the $S_i$ and
of degree $m$ in the $T_j$, and transverse in the sense that the equations
$$ 0={\partial G\over \partial S_i}={\partial G\over\partial T_j}\eqn\bocco$$
have no solutions unless either the $S_i$ or the $T_j$ are all zero.
The potential energy of this theory is (at $\sigma=0$)
$$U(s_i,t_j)={e_1{}^2\over 2}\left(\sum_i|s_i|^2-n|p|^2-r_1\right)^2
   +{e_2{}^2\over 2}\left(\sum_j|t_j|^2-m|p|^2-r_2\right)^2
   +\sum_\alpha|F_\alpha|^2, \eqn\wocco$$
with
$$\sum_\alpha|F_\alpha|^2
= |G|^2+ |p|^2\left(\sum_i\left|{\partial G\over\partial
 s_i}\right|^2+\sum_j\left|{\partial G\over \partial t_j}\right|^2\right).
\eqn\licco$$

Now, let us analyze the vacuum structure.  Transversality of $G$ means
that \licco\ vanishes precisely if either (i) $p=0$ and $G=0$; or (ii)
$s_i=0,i=1\dots n$, with suitable values of other fields;
or (iii) $t_j=0,j=1\dots m$, with suitable values of other fields.
These three choices
correspond to three phases of the theory, which we will call phase
I, phase II, and phase III.

Which phase is realized depends on the values of the $r_i$.  Upon
setting \wocco\ to zero, one finds that phase I is realized for
$r_1,r_2\geq 0$; phase II for $r_1\leq 0$, $nr_2-mr_1\geq 0$;
and phase III for $r_2\leq 0$, $nr_2-mr_1\leq 0$.  The phase
boundaries are thus the half-lines $r_1=0, r_2\geq 0$, $r_2=0,
r_1\geq 0$, and $r_1,r_2\leq 0$, $nr_2-mr_1=0$ in the $r_1-r_2$ plane.
These phase boundaries divide the plane into regions of phase
I, II, or III.

Phase I is the familiar Calabi-Yau phase.  In this phase, the expectation
values of $s_i$ and $t_j$ break the gauge group completely.
In the vacuum, $p=0$, $\sum_i|s_i|^2=r_1$, $\sum_j|t_j|^2=r_2$.
Upon dividing by the gauge group $U(1)\times U(1)$, the $s_i$ and
$t_j$ determine a point in $\CP^{n-1}\times \CP^{m-1}$ with Kahler classes
proportional to $r_1,r_2$.    Since
the vanishing of \licco\ requires also $G=0$ (and all modes not
tangent to the solution space of $G=0$ are massive), the low energy
theory is a sigma model of the hypersurface $G=0$ in $\CP^{n-1}
\times \CP^{m-1}$.  Because $G$ is of bidegree $(n,m)$, this
is actually a Calabi-Yau hypersurface.

Now let us consider phase II.  In this region, $p$ and $t_j$
have vacuum expectation values, spontaneously breaking the gauge
group to $\Z_n$.  The expectation values obey
$$|p|=\sqrt{-r_1/n}, \;\;\; \sum_j|t_j|^2=r_2-{mr_1\over n}.\eqn\ico$$
By a gauge transformation, one can assume that $p>0$.  This leaves
the residual gauge invariance
$$ t_j\to \exp(i\theta)\cdot
t_j,\;\;\; s_i\to \exp(-i(m/n)\theta)\cdot s_i.\eqn\pup$$
The value of $t_j$, after imposing the second equation in \ico\ and the
gauge invariance in \pup, determines a point in $\CP^{m-1}$.
Because $s_i$ transforms as in \pup\ under this gauge transformation,
the $s_i$ should be considered to define not a point in $\C^n$ but
a point in the fiber of an $n$ dimensional complex
vector bundle $Y$ over $\CP^{m-1}$.
(If $m/n$ is not an integer, this is not an ordinary vector bundle
but a $V$-bundle, the vector bundle analog of an orbifold.)
Setting $p$ to its vacuum expectation value, there is
an effective superpotential given by the holomorphic function
$W_{\mit eff}= \sqrt{-r_1/n}\cdot G(s_i,t_j)$
on $Y$.  Phase II is thus a peculiar hybrid of
a $\CP^{m-1}$ sigma model in the $t$ directions
and a Landau-Ginzburg orbifold in the $s$ directions
This is our first encounter with such a hybrid, but we will see that
hybrids of various kinds are ubiquitous.

Phase III can, of course, be treated in the same way and is a similar
hybrid of a $\CP^{n-1}$ sigma model and a Landau-Ginzburg orbifold.

One may wonder if it is possible to describe a Calabi-Yau hypersurface
in a product of projective spaces by a more ordinary Landau-Ginzburg
model.  The closest we can come to this is to go to the phase boundary
between phase II and phase III, setting $r_1,r_2<0$ and $mr_1-nr_2=0$.
Then there is up to gauge transformation a unique classical vacuum
with $s_i=t_j=0$, $p=\sqrt{-r_1/n}=\sqrt{-r_2/m}$.  The vacuum
expectation value of $p$ breaks the gauge group to $H=U(1)\times\Z_d$,
where $d$ is the greatest common divisor of $n$ and $m$.  The
low energy theory is a {\it gauged} Landau-Ginzburg model, with
gauge group $H$.

As an example, take $n=m=3$, with $G(S_i,T_j)$ homogeneous of degree $(3,3)$.
On the boundary between phase II and phase III, the unbroken gauge group
is $H=U(1)\times \Z_3$; $U(1)$ acts as $S_i\to \exp(i\theta)\cdot S_i$,
$T_j\to \exp(-i\theta)\cdot T_j$, and $\Z_3$ multiplies the $S_i$ by a cube
root of unity while leaving $T_j$ invariant.
The low energy theory is described by the effective action
$$L_{\mit eff}=L_{\mit gauge}+\int \d^2y\d^4\theta\left(\b S_ie^{2V} S_i
+\b T_j e^{-2V} T_j -r V\right) -\int\d^2y\left(\int\d^2\theta <p>G
+{\mit h.c.}\right), \eqn\umco$$
where the parameter $r$ measures the distance from the II-III phase
boundary.  Precisely for $r=0$,
the model has an isolated classical vacuum at the origin and unbroken
gauge invariance.  The model is then a gauged Landau-Ginzburg model
at low energies.

Gauged Landau-Ginzburg models can be analyzed like ordinary ones.
For ordinary Landau-Ginzburg models,
the contribution of the untwisted sectors to the $(c,c)$
chiral ring (the chiral ring of the $B$ model) is the ring of
polynomials $A(S_i,T_j)$ in the chiral fields modulo the usual
relations generated by $\partial G/\partial S_i$, $\partial G/\partial T_j$.
For a Landau-Ginzburg orbifold, one requires that $A$ be invariant
under the appropriate discrete symmetry.  For a gauged Landau-Ginzburg
model, $A$ must be gauge invariant.  (As we learned in \S3.1, orbifolds
are simply a special case of gauge theories with finite gauge group.)
In the present case,
gauge invariance can be imposed as follows: a $U(1)$ invariant homogenous
polynomial in $S_i$ and $T_j$
must be of equal degree in $S_i$ and $T_j$, while $\Z_3$
invariance says that this degree must be divisible by three.
The chiral ring thus consists of polynomials of bidegree $(0,0)$, $(3,3)$,
$(6,6)$, and $(9,9)$.
It seems unlikely that this model can be described by an ordinary
(ungauged) Landau-Ginzburg model.  Without the requirement of continuous
gauge invariance, the $(c,c)$ chiral ring would be infinite dimensional.

\subsection{More General Toric Varieties}

A generalization is to consider a model with $N$ chiral superfields
$S_1,\dots,S_N$, and one more chiral superfield $P$, and with
gauge group $H=U(1)^d$.  In the previous example,
$N=n+m$ and $d=2$.  One can think of the $S_i$ as linear functions
on $V=\C^N$.    We suppose that the $j^{\mit th}$ copy of $U(1)$ acts
by $S_i\to \exp(i\theta q_{ij})\cdot S_i$ with integers $q_{ij}$,
and $P\to \exp(-i\theta q_{j})P$, with $q_j=\sum_iq_{ij}$.
We pick the superpotential to be $W=P G(S_1,\dots,S_N)$ where $G$ has charge
$q_j$ under the $j^{\mit th}$ copy of $U(1)$.
The potential energy is
$$U(s_1,\dots,s_n,p)=\sum_j{1\over 2e_j{}^2}D_j{}^2+\sum_\alpha |F_\alpha|^2
 \eqn\jdjd$$
with
$$ D_j=-e_j{}^2\left(\sum_i q_{ij}|s_i|^2-q_j|p|^2-r_j\right)\eqn\oodod$$
and
$$\sum_\alpha|F_\alpha|^2=|G|^2+|p|^2\sum_i\left|{\partial G\over \partial
s_i}\right|^2. \eqn\opd$$

If one temporarily restricts to $p=0$, then setting the $D_j$ to 0 and dividing
by $H$, one obtains a so-called symplectic quotient of $V=\C^N$ by $H$,
as discussed in \S4.
This quotient, call it $Z$, is of real dimension $2(N-d)$, and
obviously admits an action of a group
$F=U(1)^{N-d}$ (the phase rotations of the $s_i$ that have not been
included in $H$).
Actually (once the $r_j$ are picked), $Z$ is naturally
a Kahler manifold -- this is obvious in the formalism
of $N=2$ supersymmetry; mathematically, the complex structure on $Z$ can
be constructed using the relation between symplectic and holomorphic
quotients, as sketched in \S4.
Using the complex structure on $Z$,
the $F$ action on $Z$ can be analytically
continued to an action of $F_\C=(\C^*)^{N-d}$.
\foot{The $F_\C$ action on $Z=V\catquot H$ can be constructed concretely
as follows.  The $F$ action on $V$ is generated by transformations
$s_i\to \lambda^{e_i}s_i$ with $\lambda\in U(1)$ and certain
exponents $e_i$.  By permitting $\lambda\in \C^*$, one extends
the $F$ action on $V$ to an $F_\C $ action; as this commutes with
$H$, it descends to an $F_\C$ action on $V/H_\C$, which as we know
from \S4 coincides with $V\catquot H$ on a dense open set.}

\REF\aubin{M. Aubin, {\it The Topology Of Torus Action On Symplectic
Manifolds} (Birkhauser, 1991).}
\REF\cox{D. A. Cox, ``The Homogeneous Coordinate Ring Of A Toric
Variety'' (preprint, Amherst College Mathematics Department, 1992).}
$Z$ and $F_\C$ have
the same complex dimension $N-d$ and $F_\C$ acts freely on a dense
open set in $Z$; $Z$ is therefore a so-called toric variety.
Under some restriction on the $q_{ij}$, $Z$ is compact.  This is true
if for some $j$ the $q_{ij}$ are all positive.  (It is true more generally
if for some positive numbers $m_j$, one has $\sum_jq_{ij}m_j>0$ for all $i$;
this however is not a very essential generalization as one can reduce
to the previous case by an automorphism of the gauge group $U(1)^d$.)
Conversely, every compact toric variety can be obtained in this way [\aubin,
\cox].

The equation $G=0$ defines a hypersurface $X$ in $Z$.
As a hypersurface in a toric variety, $X$ is called a toric hypersurface.
If we pick $G$ to be transverse in the sense that the equations
$$G={\partial G\over\partial S_i}=0\eqn\uglyl$$
are satisfied only for all $S_i=0$, then this hypersurface is smooth
away from singularities of $Z$; we henceforth consider only this case.

With $Z$ and $X$ as above and suitable $r_j$ (for instance $r_j>0$
precisely for some value of $j$ for which $q_{j}>0$),
the condition $U=0$ forces the $s_i$ to not all vanish.
Then setting $U$ to zero
(and using the transversality of $G$), $p$ and $G$ must vanish so
the locus of classical ground states up to gauge transformation
is precisely the hypersurface $X\subset Z$.
The low energy theory is a sigma model with this target space.

\REF\batyrev{V. Batyrev, ``Variations Of The Mixed Hodge Structure
Of Affine Hypersurfaces In Algebraic Tori'' (preprint, 1992).}
On the other hand, by varying the $r_j$, the system will undergo many
phase transitions.  Some of these phase transitions will involve
changes in the topology of $X$, preserving the interpretation of the
model as a Calabi-Yau sigma model (at low energies), while changing
the topology of space-time.  The local behavior involved in such topology
change will be discussed in \S5.5.
Other phases of the system will be hybrids
of sigma models and Landau-Ginzburg models; one type of hybrid
appeared in the example treated at the beginning of this subsection.
Finally,
if we set $r_j=-rq_j$ (with some fixed $r$), then
there is a unique classical vacuum up to gauge transformation; it has
$|p|^2=r$, $s_i=0$.  In this case, we get a description as
a gauged Landau-Ginzburg model, generalizing the case treated above.
The gauge group is $U(1)^{d-1}\times \Z_n$, for some $n$.
It is worth noting that Batyrev [\batyrev] gave
a kind of Landau-Ginzburg description of the cohomology of a general toric
hypersurface;  the facts we have just explained presumably give
a field theoretic setting for his construction.

\subsection{A General Comment}

A smooth toric variety $Z$ of dimension $N-d$, with $H^2(Z,{\bf R})$
being $d$ dimensional, can always be
obtained as the symplectic quotient of $\C^N$ by $U(1)^d$.
The symplectic quotient $\C^N\catquot U(1)^d$ depends on $d$
parameters $r_1,\dots, r_d$, the constant terms that can be
added to the $D$ functions.
In keeping with the Duistermaat-Heckman theorem, starting from
a fixed symplectic structure on $\C^N$, the natural induced
symplectic structure on $Z=\C^N\catquot U(1)^d$ varies linearly with the
$r_i$.  In other words, in some basis
$e_i$ of $H^2(Z,{\bf R})$, the Kahler class of $Z=\C^N\catquot U(1)^d$
is $[\omega]=\sum_ir_ie_i$.

Now consider the Calabi-Yau hypersurface $X\subset Z$. $H^2(X,{\bf R})$
has some generators that can be found by restricting the cohomology
of $Z$ to $X$.  The restriction of $H^2(Z,{\bf R})$ may however
give only part of $H^2(X,{\bf R})$ (as in the
example studied in [\amg]).  In this case, the most general Kahler
metric on $X$ is not conveniently obtained by embedding in $Z$,
and the $r_i$ are only a subset of natural linear coordinates
parametrizing the Kahler class of $X$.  When this occurs,
the phase diagram we have constructed is not the full phase diagram
of sigma models with target space $X$, but its restriction to the
subspace spanned by the $r_i$.

\section{Hypersurfaces in Grassmannians}

\REF\hubsch{T. Hubsch,  {\it Calabi-Yau Manifolds: A Bestiary For
Physicists} (World Scientific, 1992).}
Here we will consider a simple application of similar reasoning
applied to a model with a non-abelian gauge group.
In fact, we will find a correspondence between sigma models
with target space a Calabi-Yau hypersurface in a Grassmannian and gauged
Landau-Ginzburg models, this time with a non-abelian gauge group.
(Calabi-Yau hypersurfaces in Grassmannians are discussed in
Hubsch's book [\hubsch].)
In contrast to our previous experience, in this example the occurrence
of a gauged Landau-Ginzburg model will be stable, not limited
to special values of the parameters, and therefore perhaps more interesting.
Further generalizations, using more complicated gauge groups or
gauge groups in other representations, should be obvious.

To begin with, we need a convenient realization of a sigma model
with target space a Grassmannian, say the Grassmannian of $k$ planes
in complex $n$ space.  To achieve this, we start with $kn$ chiral
superfields $S^i{}_\lambda$, $i=1\dots k$, $\lambda=1\dots n$.  We think
of the $S^i{}_\lambda$ as matrix elements of a $k\times n$ matrix $S$; $\b S$
will denote its adjoint.  The group $G=U(k)$ acts on the $S$'s by
$$ S^i{}_\lambda\to M^i{}_{i'}S^{i'}{}_\lambda, \eqn\toogoo$$
for $M^i{}_{i'}\in U(k)$.  To write a model with gauge group $U(k)$,
we introduce a $k\times k$ hermitian matrix $V$ of vector superfields.
We take the Lagrangian to be
the gauge and matter kinetic energy plus a Fayet-Iliopoulos term
for the central factor $U(1)\subset U(k)$.
The potential energy is (at $\sigma=0$)
$$U(s^i{}_\lambda)= {1\over 2e^2}\Tr D^2\eqn\buxu$$
where
$$-{ D\over e^2}= s\b s - r \eqn\luxu$$
(that is, $-D/e^2$ is a $k\times  k$ hermitian matrix with
matrix elements
$-(D/e^2)^i{}_{i'}
=\sum_\lambda s^i{}_\lambda\b s^\lambda{}_{i'}-r\delta^i{}_{i'}$).
Vanishing of $D$ can be given the following interpretation.
Consider the $s^i{}_\lambda$ for fixed $i$ as defining a vector
$w^{(i)}$ in an $n$ dimensional vector space $W\cong \C^n$.
(If you wish, pick a fixed basis $e^1\dots e^n$ of $W$ and set
$w^{(i)}=\sum_\lambda s^i{}_\lambda e^\lambda$.)
Then the equation $D=0$ asserts
\foot{For $r>0$. For $r<0$, the renormalization effect
studied in \S3.2 is important in understanding the model; the vacuum
states all lie at large $\sigma$.}
that the $w^{(i)}/\sqrt r$ are orthonormal.  These vectors therefore span
a $k$ dimensional subspace $F$ of $W$.  Upon dividing by the gauge group
$U(k)$, the gauge invariant information contained in the expectation
values of the $s^i{}_\lambda$ is merely the choice of $F$.  The space
of all $F$'s is by definition the Grassmannian ${\bf G}(k,n)$ of $k$-planes
in complex $n$ space.
At low energies, the model is thus simply a sigma
model with this target space.  In mathematical terms, using the language
of \S4, the above construction amounts to obtaining ${\bf G}(k,n)$
as the symplectic quotient of $\C^{kn}$ by $U(k)$.
\foot{Alternatively, the Grassmannian can be realized as a holomorphic
quotient as follows.  By allowing $M\in GL(k,\C)$, \toogoo\ defines
a $GL(k,\C)$ action on $\C^{kn}$; the Grassmannian is
the quotient by $GL(k,\C)$ of a dense open set in $\C^{kn}$ consisting
of good orbits that contain zeros of the moment map.  These are precisely
the orbits on which the vectors $w^{(i)}=s^i{}_\lambda e^\lambda$ are linearly
independent.}

To obtain a Calabi-Yau hypersurface in this Grassmannian,
we introduce
one more complex superfield $P$,
transforming under \toogoo\ as $P\to (\det M)^{-n}P$.
One then takes the superpotential to be $W=P G(S^i{}_\lambda)$, where
$G$ is a polynomial that transforms as $G\to (\det M)^{n}G$.
The formula for $D$ is modified to
$$-{ D\over e^2}= s\b s - kn|p|^2-r \eqn\vluxu$$
The additional terms that appear in the scalar potential are
$$ \sum_\alpha|F_\alpha|^2=|G|^2+|p|^2\sum_{i,\lambda}\left|
{\partial G\over\partial s^i{}_\lambda}\right|^2.  \eqn\oci$$

For $r>>0$, the model can be analyzed in a now familiar fashion.
Assuming that $G$ is chosen so that the hypersurface $X\subset {\bf G}(k,n)$
given by $G=0$ is smooth, vanishing of the potential
implies that $p=G=0$; the low energy theory is a sigma model with
target space $X$.  For $r<<0$, under the same assumptions,
the model has a unique classical vacuum (up to gauge transformation)
with $<p>=\sqrt{-r/kn}$, $<s^i{}_\lambda>=0$.  The expectation value
of $p$ breaks the gauge group from $U(k)$ to the subgroup $H$ of
$U(k)$ consisting of matrices whose determinant is an $n^{\mit th}$
root of 1.  ($H$ is an extension of $SU(k)$ by $\Z_n$.)  The model
at low energies is a gauged Landau-Ginzburg model with
massless chiral superfields $S^i{}_\lambda$, effective superpotential
$W_{{\mit eff}}=<p>G(S^i{}_\lambda)$, and gauge group $H$.

\subsection{An Example}

Here is a convenient and amusing example (see [\hubsch], p. 101).
One of the simplest examples of a Calabi-Yau manifold $X$ is an
intersection $G_1=G_2=0$ in $\CP^5$, with $G_1$ and $G_2$
being respectively polynomials homogeneous of degree two and of degree
four in homogeneous
coordinates $T_1,\dots, T_6$ of $\CP^5$.
It does not seem that this model has an ordinary
Landau-Ginzburg realization.  However, it has a gauged
Landau-Ginzburg phase, since (for a smooth quadric $G_1$) the
locus $G_1=0$ in $\CP^5$ is equivalent to a copy of ${\bf G}(2,4)$.
Hence, $X$ can be realized as a hypersurface in ${\bf G}(2,4)$, and studied
as above.

This may be seen explicitly as follows.  Instead of thinking of
the $T_i$ as a six component ``column vector,'' arrange them
as components of a $4\times 4$ antisymmetric tensor $T_{\lambda\beta}=-
T_{\beta\lambda},
\,\,\lambda,\beta=1\dots 4$.
There is no invariant information in the choice of the quadratic polynomial
$G_1$, assuming it is non-degenerate,
since all non-degenerate quadrics are equivalent up to a linear
change of coordinates.
A convenient choice of $G_1$
is
$$ G_1=\epsilon^{\alpha\beta\gamma\delta}T_{\alpha\beta}T_{\gamma\delta}
             \eqn\osos$$
with $\epsilon^{\alpha\beta\gamma\delta} $ the Levi-Civita tensor.
The equation $G_1=0$ can be solved in terms of eight complex variables
$S^i{}_\lambda$, $i=1,2$, $\lambda=1\dots 4$, by
$$T_{\lambda\delta}=\epsilon_{ij}S^i{}_{\lambda}S^j{}_{\delta} \eqn\boso$$
with $\epsilon_{ij}$ again the Levi-Civita tensor.
Two $S^i{}_\lambda$'s give $T$'s that are proportional
(and so define the same point in $\CP^5$) precisely if they
differ by $S^i{}_\lambda\to
M^i{}_{i'}S^{i'}{}_\lambda$, $M$ being in $GL(2,\C)$.  Since
this equivalence relation on the $S$'s is the one that leads to
${\bf G}(2,4)$, this completes the explanation of the isomorphism
of the quadric $G_1=0$ in $\CP^5$ with ${\bf G}(2,4)$.

We want to study the Calabi-Yau manifold $X$ given by the equations
$G_1=G_2=0$ in $\CP^5$, with $G_1(T_\lambda)$ as above and $G_2$ a quartic
polynomial in the $T$'s.  Under the substitution
\boso, $G_2(T)$ becomes an eighth order, $SL(2,\C)$
invariant polynomial $\widetilde G(S)$.
$X$ is equivalent to the hypersurface $\widetilde G(S)=0$ in ${\bf G}(2,4)$.
So according to our
general discussion of hypersurfaces in Grassmannians, the sigma
model  with target space $X$
can be studied by studying the $U(2)$ gauge theory with
chiral superfields $S^i{}_\lambda$ and superpotential $W=P\widetilde G(S)$.
Its phases can be found as at the beginning of this section
and are the Calabi-Yau phase and a gauged Landau-Ginzburg phase.

\section{Intersections Of Hypersurfaces}

Now we will briefly discuss the application of these ideas
to intersections of hypersurfaces.

To avoid cluttering the notation, we will consider only the
simplest case of an intersection of hypersurfaces in
a ordinary
projective space, say $\CP^{n-1}$ with projective coordinates $s_1,\dots
, s_n$.  We consider hypersurfaces $H_a$, $a=1\dots k$ defined
by the vanishing of homogeneous polynomials $G_a$ of degree $q_a$.
The intersection $X$ of the $H_a$ is a Calabi-Yau manifold if
$\sum_a q_a = n$.  The condition that $X$ is smooth is that
the $H_a$ intersect transversely in the sense that
for any complex numbers $p_a$, not all zero, the equations
$$G_a= \sum_ap_a{\partial G_a\over \partial s_i}=0 \eqn\jujo$$
have a common solution only for $s_1=\dots = s_n=0$.

To describe the hypersurface $X$ along the general lines of the present
paper, we consider a $U(1)$ gauge theory with chiral superfields
$S_i$ of charge 1 and $P_a$ of charge $-q_a$.
We take the superpotential
to be $W=\sum_aP_a G_a(S_1,\dots,S_n)$.  The ordinary potential is then
$$U={1\over 2e^2}D^2+\sum_\alpha |F_\alpha|^2\eqn\nurkok$$
with
$$D=-e^2\left(\sum_i|s_i|^2 -\sum_a q_a|p_a|^2 -r  \right)\eqn\jombo$$
and
$$\sum_\alpha|F_\alpha|^2 = \sum_a|G_a|^2+\sum_i\left|\sum_ap_a{\partial G_a
\over \partial s_i}\right|^2. \eqn\combo$$
There are two phases.  For $r>0$,
one uses the transversality condition \jujo\ to show (as we have done
earlier in similar problems) that the space of classical ground states
is the variety $X$.  For $r<0$, the $p$'s have an expectation value,
and the $s$'s do not.
The space of ground states is then
a weighted projective space $W={\bf WCP}^{k-1}
_{a_1,\dots,a_n}$, and the low energy theory is a hybrid Landau-Ginzburg/sigma
model on a vector bundle over $W$.

Intersections of hypersurfaces in more general toric varieties
(or Grassmannians, etc.) can be treated by obvious extensions of these
remarks.  In general, one will get many phases, of all the types
we have seen (Calabi-Yau, Landau-Ginzburg, gauged Landau-Ginzburg,
and various kinds of hybrid).  One may wonder if there are
any cases in which one gets an ordinary Landau-Ginzburg
phase as opposed to a sigma model/L.G. hybrid.  This occurs only in special
cases, some of which are of phenomenological interest.  I will next
describe such a case.

\subsection{Landau-Ginzburg Models For Intersections Of Hypersurfaces}

In $\CP^{n-1}\times \CP^{m-1}$, with homogeneous coordinates $s_1,\dots,
s_n$ and $t_1,\dots , t_m$ for the two factors, we will consider
the variety $X$ defined by $G_1=G_2=0$,
the $G$'s  being bi-homogeneous polynomials in $s_i$ and in $t_j$.
If $G_a,a=1,2$
is bi-homogeneous in $s_i,t_j$ of degree $(q_a,q'_a)$, then $X$ is
a Calabi-Yau manifold if $q_1+q_2=n,q'_1+q'_2=m$.
The condition that
$X$ is smooth is that for any complex numbers $p_a$
not both zero, the equations
$$0  = G_1=G_2=\sum_ap_a{\partial G_a\over\partial s_i}=\sum_ap_a{\partial
 G_a\over \partial t_j}\eqn\hooom$$
have no common solution unless the $s_i$ or the $t_j$ are all zero.
(Values of $(s_i,t_j)$ with all $s_i$ or all $t_j$ zero
do not determine a point in
$\CP^{n-1}\times \CP^{m-1}$.)  This transversality condition,
however, is not sufficient
to lead in our usual construction
to a Landau-Ginzburg model; it
will lead to a hybrid Landau-Ginzburg/sigma model similar to some we have
seen above.

To construct a model with the sigma model of $X$ as one of its phases,
we consider a $U(1)\times U(1)$ gauge theory, with chiral fields
$S_i,\,i=1\dots n$ of charge $(1,0)$, $T_j,\,j=1\dots m$ of charge
$(0,1)$, and two more fields $P_a$, $a=1,2$, of charge $(-q_a,-q'_a)$.
We pick the superpotential to be $W=\sum_aP_aG_a(S_i,T_j)$.  The relevant
parts of the classical potential are
$$\sum_a{1\over 2e_a{}^2}D_a{}^2={e_1{}^2\over 2}
\left(\sum_i |s_i|^2-\sum_aq_a|p_a|^2-r_1\right)^2
+{e_2{}^2\over 2}
\left(\sum_j |t_j|^2-\sum_aq'_a|p_a|^2-r_2\right)^2   \eqn\ropo$$
and
$$\sum_\alpha |F_\alpha|^2=|G_1|^2+|G_2|^2
+\sum_i\left|\sum_ap_a{\partial G_a\over\partial s_i}\right|^2
+\sum_j\left|\sum_ap_a{\partial G_a\over\partial t_j}\right|^2 .\eqn\cobo$$
Setting all this to zero, one finds for $r_1,r_2>>0$ a phase which
describes at low energies a sigma model with target space $X$.
In general there is a perhaps interesting phase diagram with several phases,
but none of these describes a Landau-Ginzburg model.  For instance,
for $r_1,r_2<<0$, setting $D_a$ to zero requires $p_1,p_2$ to not both
vanish.
The vanishing of the $F_\alpha$ is then equivalent to \hooom, and
the transversality condition asserts that this requires that
the $s_i$ are all zero or the $t_j$ are all zero -- but not necessarily both.
If the $s_i$ are all zero, the $t_j$ are not uniquely determined,
and vice-versa.
One therefore has not an isolated classical vacuum, leading to a
Landau-Ginzburg model, but a family of vacua with additional massless
particles, corresponding to a hybrid model.  This particular
hybrid model actually has a
reducible target space (one branch with $s_i=0$ and one
with $t_j=0$) and a singularity (other than the usual orbifold
singularities) at the intersection $s_i=t_j=0$.

A Landau-Ginzburg model arises only if $n,m$ and the $q_a$, $q'_a$
are such that, for suitable values of $r_1$ and $r_2$, vanishing
of the classical potential forces the $s_i$ {\it and} $t_j$ all to vanish.
This occurs only in very special cases.\foot{
Apparently the relevant cases are precisely $n=m$, $(q_1,q_1{}')
=(n-1,1)$, $(q_2,q_2{}')=(1,n-1)$; or alternatively $n\geq m$,
$(q_1,q_1{}')=(n-1,0)$, $(q_2,q_2{}')=(1,m)$.}  Some of these
pecial cases are of phenomenological interest.  For instance,
take $n=4$, $m=3$, $(q_1,q_1{}')=(3,0)$, $(q_2,q_2{}')=
(1,3)$.  We take
$$G_1=\sum_{i=1}^4 S_i{}^3,\;\;\;\;G_2=\sum_{j=1}^3S_iT_i{}^3.\eqn\mcco$$
We have
$$\sum_a{1\over 2e_a{}^2}D_a{}^2={e_1{}^2\over 2}
\left(\sum_i |s_i|^2-3|p_1|^2-|p_2|^2-r_1\right)^2
+{e_2{}^2\over 2}
\left(\sum_j |t_j|^2-3|p_2|^2-r_2\right)^2   \eqn\ropop$$
We consider the region $r_1<<r_2<<0$.  Vanishing of \ropop\
requires that $p_2\not=0$ and that either (a) $p_1\not=0$, or
(b) $p_1=0$, and the $t_j$ are not all zero.  Case (b) is incompatible
with \hooom, which immediately implies that the $t_j$ are all zero
if $p_1=0$, $p_2\not=0$.  In case (a), \hooom\
is easily seen to imply that $s_i=t_j=0$ for all $i$ and $j$.
In this case, $|p_1|$ and $|p_2|$ are uniquely determined
by \ropop.  So for this range of $r_1$ and $r_2$, the model
has a unique classical vacuum up to gauge transformation and
reduces at low energies to a Landau-Ginzburg orbifold.
The model is actually a $\Z_9$ orbifold, as the vacuum expectation
values of $p_1$ and $p_2$ break the gauge group to $\Z_9$.
The effective superpotential for the
massless superfields $S_i,T_j$ is (setting the $<p_a>$ to 1 by rescaling
the variables)
$$W_{\mit eff}=\sum_aG_a=\sum_{j=1}^3\left(S_j{}^3+S_jT_j{}^3\right)
+S_4{}^3.   \eqn\ikco$$

\REF\schimmrigkk{R. Schimmrigk, ``A New Construction Of A Three Generation
Calabi-Yau Manifold,'' Phys. Lett. {\bf 193B} (1987) 175.}
\REF\gepner{D. Gepner, ``Exactly Solvable String Compactification
On Manifolds of $SU(N)$ Holonomy,'' Phys. Lett. {\bf 199B} (1987) 380;
``String Theory On Calabi-Yau Manifolds: The Three Generations Case''
PUPT-88/0085 (unpublished); ``Space-Time Supersymmetry In Compactified
String Theory And Superconformal Models,'' Nucl. Phys. {\bf B296} (1988) 757.}
This model is of interest for two reasons.  First of all, a suitable
quotient of the variety $X$ by a finite group gives one of the first
known and simplest three
generation models, constructed originally by Schimmrigk [\schimmrigkk].
Second, the Landau-Ginzburg description shows that the model is exactly
soluble at a particular point in its moduli space.  Indeed,
$S^3+ST^3$ is the superpotential of the minimal model with $ E_7$
modular invariant, while $S^3$ is that of the minimal model with
$A_1$ modular invariant, so the model
is a tensor product of three $E_7$ minimal models
and one $A_1$ model.  It is in fact one of Gepner's original models
[\gepner].

\section{Change Of Topology}
It has been a long-standing and fascinating question whether in string
theory physical processes can occur in which the topology of space-time
changes.  (Examples have been known, involving duality or mirror symmetry,
of non-classical equivalences between
different topologies, but that is a somewhat different question.)
Recently, Aspinwall, Greene, and Morrison [\amg] undertook to use mirror
symmetry to show that the answer is affirmative, at least in one special
situation.  Though their starting point is quite different, their analysis
ultimately involves
the same phase diagrams that we have been examining in this paper.

These phase diagrams can be very complicated in general
and -- as shown in detail in [\amg] in a particular example --
can contain a variety of Calabi-Yau phases, with different smooth target
spaces.  Upon varying the Fayet-Iliopoulos terms,
transitions occur between the different target space-times.
By arguments of \S3.2, the transitions are continuous if the $\theta$ angles
are generic.  In string theory, the Fayet-Iliopoulos parameters (like all
operators in the world-sheet Lagrangian) are interpreted as expectation
values of some fields in space-time; they are dynamical variables, free
to change in time.  This gives a mechanism for physical topology
change.

The transitions that arise this way are transitions between
topologically distinct but birationally equivalent Calabi-Yau manifolds.
The reasons for this were sketched in \S4; in brief, the various
symplectic quotients of a given target space that one can construct
by varying the Fayet-Iliopoulos parameters can all be identified
generically (on a dense open set) with a fixed complex quotient of the same
target.

While a global situation has been analyzed in [\amg],
my discussion here will be more modest in scope, focussing on
the local mechanism of topology change.  In other words,
we will analyze how, by varying the constant term in the $D$ function,
we obtain a change of topology in a local region of space-time.
The reader should think of the space-time we will consider as being
embedded as part of a compact Calabi-Yau manifold, whose topology will
change in the process we will consider.

\REF\moritheory{J. Koll\'ar, ``The Structure Of Algebraic Threefolds:
An Introduction To Mori's Program,'' Bull. Am. Math. Soc. {\bf 17} (1987)
211.}
The example that we will consider is in complex dimension three and
is not as special as it may appear.  It is the generic ``flop'' (or
birational transformation not changing $c_1$) considered in Mori theory
(of birational classification of complex manifolds;
for a review see [\moritheory])
in complex dimension three.  It is possible that by introducing
enough parameters, any birational transformation between three-dimensional
Calabi-Yau manifolds can be constructed as a succession of transformations
each locally isomorphic to the one we will analyze.

We consider a $U(1)$ gauge theory with two chiral superfields
$A_i$, $i=1,2$ of charge 1 and two chiral superfields $B_j$, $j=1,2$,
of charge $-1$.  We can think of them as coordinates on $V\cong \C^4$.
We take the superpotential to be zero.  The ordinary
potential energy for the bosonic components is (at $\sigma=0$)
$$U(a_i,b_j)={e^2\over 2}\left(\sum_i|a_i|^2-\sum_j|b_j|^2-r\right)^2.
 \eqn\cconcn$$
The $U(1)$ action on $V$ is
$$\eqalign{a_i & \to \lambda a_i \cr
           b_j & \to \lambda^{-1}b_j \cr}\eqn\kiop$$
with $\lambda\in U(1)$; it can be extended to a $\C^*$ action
by permitting $\lambda\in \C^*$.

The low energy effective space-time is obtained as usual by setting $U=0$ and
dividing by $U(1)$; it is in other words $V\catquot U(1)$, the symplectic
quotient of $V$ by $U(1)$ in the language of \S4.
There are really three possible symplectic quotients,
corresponding to $r>0$, $r=0$, and $r<0$.  We will call them $Z_+$,
$Z_0$, and $Z_-$.

$Z_+$, $Z_0$, and $Z_-$ are all birationally isomorphic in their
natural complex structures according to the arguments of \S4.
In fact, let $\widetilde V$ be the region of $V$ in which the $a_i$
are not both zero, and the $b_j$ are not both zero.  Then by precisely
the arguments of \S4, the symplectic quotient $\widetilde V\catquot U(1)$
can be naturally identified -- for any $r$ -- with the holomorphic
quotient $\widetilde V/\C^*$.  $Z_+$, $Z_0$, and $Z_-$ are partial
compactifications of $\widetilde V/\C^*$ obtained by including
contributions of some of the bad $\C^*$ orbits on which $a_1=a_2=0$
or $b_1=b_1=0$.    In fact, these bad orbits are of three types:
the origin $O$ with $a_i=b_j=0$; the set $V_1$ of orbits with $b_j=0$
but the $a_i$ not both zero; and the set $V_2$ of orbits with $a_i=0$
but the $b_j$ not both zero.

By including the appropriate bad orbits, $Z_+$, $Z_0$, and $Z_-$
can be described as follows:

(i) For $r>0$, the $\C^*$ orbits that do not contain zeros of $D$ are
the orbits in $O$ and $V_2$.  Other $\C^*$ orbits (whether in
$\widetilde V$ or $V_1$)
each contribute precisely one point to the symplectic quotient $Z_+$.
Hence $Z_+=(\widetilde V\cup V_1)/\C^*$.  Here $\widetilde V\cup V_1$
is precisely the region in $V$ in which the $a_i$ are not both zero.
The values of the $a_i$, up to scaling by $\C^*$,
determine a point in a copy of $\CP^1$
which we will call $\CP^1_a$.  $Z_+$ is fibered over  $\CP^1_a$ by forgetting
the values of the $b_j$; since the values of the $b_j$ are arbitrary,
the fiber is a copy of $\C^2$.
The zero section of $Z_+\to \CP^1_a$, that is the locus
$b_1=b_2=0$, is
a genus zero holomorphic curve, in fact an embedding $\CP^1_a\subset Z_+$.
We henceforth identify $\CP^1_a$ with its image under this embedding.
For $b_j=0$, vanishing of $D$ gives $\sum_i|a_i|^2=r$, so the Kahler
form of $\CP^1_a$ is proportional to $r$.

(ii) For $r=0$, the only bad $\C^*$ orbit that contributes is $O$;
it contributes a single point to $Z_0$, and that point is a singularity.
We will examine the singularity more closely later.

(iii) The structure for $r<0$ is similar to that for $r>0$ with the roles
of $a_i$ and $b_j$ reversed.  In particular, $Z_-=
(\widetilde V\cup V_2)/\C^*$.  Here $\widetilde V\cup V_2$ is the region
of $V$ in which the $b_j$ are not both zero.  The values of the $b_j$
determine a point
in a copy of $\CP^1$ that we will call $\CP^1_b$.  $Z_-$ is fibered
over $\CP^1_b$ with fiber a copy of $\C^2$ parametrized by the valued of
the $a_i$.  The zero section $a_1=a_2=0$ of $Z_-\to \CP^1_b$ gives
a genus zero curve in $Z_-$ which is an embedding $\CP^1_b\subset Z_-$.
This curve -- which we identify with $\CP^1_b$ -- has Kahler form proportional
to $-r$.

$Z_+$, $Z_0$, and $Z_-$ are all non-compact Calabi-Yau manifolds.
Indeed, the holomorphic four-form $\Theta= \d a_1\wedge \d a_2\wedge \d b_1
\wedge
\d b_2$ on $V$ is $\C^*$ invariant (because $\sum_iQ_i=0$); contracting
it with the vector field generating the $\C^*$ action gives an
everywhere non-zero holomorphic three form whose restriction to the
appropriate region of $V$ is the pullback of a holomorphic volume form
on $Z_+$, $Z_0$, or $Z_-$.

In passing from $r>0$ to $r<0$, a change in topology occurs.
$\CP^1_a$ shrinks to zero size as $r\to 0$ from above,
and is replaced by $\CP^1_b$ for $r<0$.
Classically the interpolation from $r>0$ to $r<0$ passes through
a singularity at $r=0$, but according to the argument of \S3.2,
there is no such singularity in the sigma model as long as the $\theta$
angle is generic.  We have achieved the promised smooth change in topology
in sigma models of (non-compact) Calabi-Yau manifolds.

It is true that (by exchanging the $a$'s and $b$'s), $Z_+$ and $Z_-$
are actually isomorphic.  However, in a global situation, with our
discussion applied to a portion of
some compact Calabi-Yau manifold, the replacement
of $Z_+$ by $Z_-$ or of $\CP^1_a$ by $\CP^1_b$
entails a change in the topology of space-time.

To give a new perspective on the model, let us now consider
more closely the nature of the singularity at the origin of the
exceptional symplectic quotient $Z_0$.

Let $x,y,z, $ and $t$ be coordinates on a copy of $\C^4$ that we will
call $W$.  The formulas
$$\eqalign{
x = & a_1b_1 \cr
y = & a_2b_2\cr
z = & a_1b_2 \cr
t = & a_2b_1 .\cr} \eqn\impigo$$
give a $\C^*$ invariant map $V\to W$.  By restricting these formulas
to $\widetilde V\cup O$ and dividing by $\C^*$,
we get a map from $Z_0\to W$.  (By starting with
$\widetilde V\cup V_1$ or $\widetilde V\cup V_2$ we would get maps
$Z_\pm \to W$.)  It is evident that the image of $Z_0$ in $W$ lies in the
affine quadric $Q$ defined by
$$ xy-zt=0. \eqn\hongo$$
In fact, the map defined in \impigo\ is an isomorphism between $Z_0$ and $Q$.

This assertion can be justified as follows:

(a) To prove that the map is surjective, we must show that if
$x,y,z$, and $t$ obey \hongo, then \impigo\ is satisfied for some values
of $a_i,b_j$.  If $x=y=z=t=0$, we take $a_i=b_j=0$.  If, say, $x\not =0$,
we pick $a_1=1$ and then
iteratively solve \impigo\ for $a_2,b_1,b_2$.

(b) To prove that the map is injective, we must show that  if $x,y,z,$ and $t$
obey \hongo, then \impigo\ determines the $a_i,b_j$ uniquely up to the
action of $\C^*$.  If $x=y=z=t=0$, then \impigo\ requires that either
the $a_i$ or the $b_j$ are both zero.  But in that case,
for $a_i$ and $b_j$ to define
a point in $Z_0$, they must all be zero, so $x=y=z=t=0$ is the image
only of the point $O\in Z_0$.  Otherwise, if say $x\not= 0$, then
\impigo\ requires $a_1\not=0$.  We can use the $\C^*$ action to set $a_1=1$,
and then \impigo\ uniquely determines $a_2,b_1,b_2$.
This completes the proof of isomorphism of $Z_0$ and $Q$.

The singularity of $Q$ at $x=y=z=t=0$ is the simplest type of
isolated singularity of a three dimensional Calabi-Yau manifold
(see [\hubsch], p. 121).
This singularity can be resolved by blowing up the origin in $Q$, but
that would ruin the Calabi-Yau condition.  There are two minimal
ways to resolve the singularity of $Q$ while preserving the Calabi-Yau
condition.  The two choices are to replace the singular point by a copy
of $\CP^1$ which should be either $\CP^1_a$ or $\CP^1_b $ as constructed
above.  Thus the two ``small resolutions'' of $Q$ as a Calabi-Yau manifold
are precisely $Z_+$ and $Z_-$.  The topology-changing transition that
we found above was a transition between these two small resolutions.

\subsection{The Instanton Sum}

To get more insight, we want to study the behavior of physical observables
in the transition from $Z_+$ to $Z_-$, or more exactly from
$X_+$ to $X_-$ where $X_\pm$ are compact Calabi-Yau manifolds
that coincide outside of a region (or perhaps finitely many
similar regions) where they look like $Z_\pm$.
The simplest physical observables to analyze
are low energy Yukawa couplings.  Yukawa couplings of massless multiplets
derived from $H^{2,1}(X_\pm)$
are independent of $r$ and so should be invariant
under the transition from $X_+$ to $X_-$.   Of more interest are the
Yukawa couplings of massless multiplets associated with $H^{1,1}(X_\pm)$.
These are determined by complicated instanton sums; we want to see
what happens to those sums in interpolating from $r>0$ to $r<0$.

Writing down the full instanton sums on $X_\pm$ would require very
detailed information about the rational curves.  Comparing the instantons
of $X_\pm$ is much easier.  All that happens in the transition from
$X_+$ to $X_-$ is that one genus zero curve, called $\CP^1_a$ above,
is lost, and replaced by another genus zero curve, above called $\CP^1_b$.
Our problem is simply to evaluate the contributions for $r>0$ from
$\CP^1_a$ and its multiple covers, and compare to the contributions
for $r<0$ from $\CP^1_b$ and its multiple covers.

First of all, if $H^{2,0}(X_\pm )=0$ (as for almost all Calabi-Yau manifolds),
then $H^{1,1}(X_\pm)$ can be identified with the group of divisors on $X_\pm$,
up to linear equivalence.  Moreover, as $X_+$ and $X_-$ differ only
in complex codimension two, divisors on $X_+$ can be naturally identified
with divisors on $X_-$.  This gives a natural isomorphism between
the groups $H^{1,1}(X_\pm)$, which we will therefore call simply $H^{1,1}(X)$.

It may appear that we need some detailed information about $H^{1,1}(X)$, but
happily that is not so.  If $E_1,E_2,$ and $E_3$ are three divisors,
the contribution of a rational curve $C$ to the corresponding Yukawa coupling
is proportional to
$$ (C,E_1)(C,E_2)(C,E_3) \eqn\gogo$$
where $(C,E)$ is the intersection number of the curve $C$ and the divisor $E$.
Therefore $\CP^1_a$ and $\CP^1_b$ and their covers contribute only
to Yukawa couplings of states associated with divisors that they meet.
This means that any divisor which, up to linear equivalence, can be
chosen not to intersect $\CP^1_a$ and $\CP^1_b$ is  not sensitive to
the interpolation from $X_+$ to $X_-$.  We therefore need not understand
the divisor classes  of $X_\pm$; it is enough to understand the divisor
classes of
$Z_\pm$ or equivalently of $Z_0$.

\REF\hartshorne{R. Hartshorne, {\it Algebraic Geometry} (Springer-Verlag,
1977).}
The divisor class group of $Z_0$ is calculated in Hartshorne's
book [\hartshorne],
example II.6.6.1 and exercise II.6.5.
The result is that any divisor on $Z_0$ (or $Z_\pm$) is a multiple
of the divisor $E$ given by $a_1=0$.  (Not being $\C^*$ invariant,
$a_1$ is a section of a line bundle rather than a function, so
the fact that $E$ is the divisor of $a_1$ does not make it trivial.)
For example, the divisor $E'$ given by $b_1=0$ obeys
$$ E+E'=0 \eqn\jucux$$
since the product $a_1b_1=x$ is a function on $Z_0$, and $E+E'$ is
the divisor of this function.

Therefore, the only Yukawa coupling that we need to evaluate is the
three point function of the multiplet associated with $E$.  To evaluate
the contributions of $\CP^1_a$ and $\CP^1_b$ to this three point coupling,
we need to know their intersection number with $E$.  Indeed,
$$(\CP^1_a,E)= 1 , \eqn\huppo$$
since $\CP^1_a$ and $E$ meet transversely in one point (represented
on $V$ by the $\C^*$ orbit $a_1=b_1=b_2=0$, $a_2\not= 0$).
Likewise $(\CP^1_b,E')=1$, and in view of \jucux, it follows that
$$(\CP^1_b,E)=-1. \eqn\uppo$$
($E$ and $\CP^1_b$ do not meet transversely, so this number cannot
be determined by just counting intersection points.)

Apart from these intersection numbers, to evaluate the instanton sums
we need to know the instanton action, which for an instanton $C$
contributes a factor
$\exp(2\pi i \tau)$, where
$-2\pi i \tau = \int_C\omega$; here $\omega$ is a two form representing
a complexified Kahler form on $X$ (its real part is the ordinary Kahler
form, and its imaginary part incorporates the
$\theta$ angle).  We therefore need $-2\pi i t_a=\int_{\CP^1_a}\omega$,
$-2\pi i t_b=\int_{\CP^1_b}\omega$.  Because every divisor is locally
equivalent to a multiple of $E$, we can assume that $\omega$ is
a multiple of the Poincar\'e dual of $E$ (plus terms that do not contribute
to $t_a$ or $t_b$).  In view of \huppo\ and \uppo\ we get therefore
the important result
$$ t_b=-t_a.      \eqn\lipop$$

\REF\dine{M. Dine, N. Seiberg, E. Witten, and X.-G. Wen, ``Non-Perturbative
Effects On The String World Sheet I,II,'' Nucl. Phys. {\bf B278} (1986) 769,
{\bf B289} (1987) 319.}
\REF\candelas{P. Candelas, P. Green, L. Parke, and X. de la Ossa, ``A Pair
Of Calabi-Yau Manifolds As An Exactly Soluble Superconformal Field Theory,''
Nucl. Phys. {\bf B359} (1991) 21.}
\REF\aspinwall{P. Aspinwall and D. B. Morrison, ``Topological Field Theory
And Rational Curves,'' Oxford preprint, to appear in Commun. Math. Phys.}
We will now compute on $X_\pm$ the Yukawa coupling of three fields
associated with the divisor $E$.  We will do the calculation as a function
of $t_a$ or equivalently $t_b$.  In the calculation on $X_+$,
the physical region is ${\rm Im}\, t_a>0$, since if the curve $\CP^1_a$
is to exist, as it does on $X_+$, it must have positive area.
The physical region for the calculation on $X_-$ is likewise ${\rm Im}\,
t_b>0$ or equivalently ${\rm Im}\,t_a<0$.  What we want to do is
to compute the instanton sum on $X_+$ for ${\rm Im}\,t_a>0$ and compare
its analytic continuation to the instanton sum on $X_-$ for ${\rm Im}\,t_a<0$.

The actual computation is not difficult.
On $X_+$, we have to take account of $\CP^1_a$ and its multiple covers.
Using the basic formula for the instanton contribution [\dine]
and the formula for the contributions of multiple covers
conjectured by Candelas et. al. [\candelas]
and justified by Aspinwall and Morrison [\aspinwall],
the contribution of $\CP^1_a$ and its multiple covers
to the Yukawa coupling is
$$\lambda_+=(\CP^1_a,E)^3\sum_{n=1}^\infty
e^{2\pi in t_a}={e^{2\pi i t_a}
\over 1-e^{2\pi i t_a}} .             \eqn\coxox$$
The $n^{th}$ term is the contribution of the $n^{th}$ cover of $\CP^1_a$;
the series converges for ${\rm Im}\,t_a>0$.

On $X_-$, we have to likewise take account of $\CP^1_b$ and its covers.
The formula analogous to \coxox\ (convergent now for ${\rm Im}\,t_a<0$)
is
$$\lambda_-=(\CP^1_b,E)^3\sum_{n=1}^\infty e^{2\pi i n t_b}
=-{e^{-2\pi i t_a}\over 1-e^{-2\pi i t_a}}. \eqn\pillo$$

One might naively think that the hypothesis of smooth continuation
from $X_+$ to $X_-$ should mean that $\lambda_+=\lambda_-$, but in fact
according to the above formulas
$$\lambda_+-\lambda_-= -1 . \eqn\willow$$
The discrepancy has a natural interpretation, however.
In addition to the instanton sums, the Yukawa couplings receive
a constant ``classical'' contribution from the intersection pairings
of $X_+$ and $X_-$.  The birationally equivalent manifolds $X_+$ and
$X_-$ have different cohomology rings and different intersection pairings.
It must be the case that the difference in intersection
pairings between $X_+$ and $X_-$ is $+1$, cancelling \willow.

This is easy to verify.  Instead of comparing the triple intersection number
$(E,E,E)$, we can just as well look at $(E,E',E')$, since restricted
to $V_\pm$, $E'=-E$.  The difference in the classical intersection
pairing between $X_+$ and $X_-$ can be measured by counting intersections
in $V_+$ and $V_-$.  We have to be careful, though, to use the same
representatives for $E,E',E'$ in $V_+$ as in $V_-$; if the divisors
are shifted by linear equivalence, the number of intersections outside
of $V_\pm$ might change.  So we represent $E$ by $a_1=0$ and
the two copies of $E'$ by $b_1=0$ and $b_2=0$.  In $V_+$,
these divisors meet transversely at the one point $a_1=b_1=b_2=0$,
so the intersection number is $+1$.  In $V_-$, $b_1$ and $b_2$
never both vanish, so the intersection number is 0.  So we get the
expected excess intersection number in $X_+$ relative to $X_-$ of $+1$.

So the Yukawa couplings on $X_-$ are analytic continuations of those
of $X_+$, as expected.
Moreover, the fact that \coxox\ and \pillo\
have a pole at $t_a=0$ as their only singularity is in accord with the
argument of \S3.2 according to which
$r$ and $\theta$ (essentially the
imaginary and real parts of $t_a$) must both vanish to get a singularity.

\chapter{EXTENSION TO $(0,2)$ MODELS}

Models with $(0,2)$ supersymmetry (that is, two right-moving and
no left-moving supersymmetries, as opposed to the $(2,2)$ models
studied above) are important phenomenologically.  In the context
of compactification of string theory to four dimensions, they lead
naturally to $SU(5)$ or $SO(10)$ rather than $E_6$ as effective
the grand unified gauge group.  These models have the reputation of being
much harder to study than $(2,2)$ models, because the analogs of
the Gepner soluble models, the Landau-Ginzburg correspondence,
etc., have not been known.  In this section we will construct the
$(0,2)$ version of the Landau-Ginzburg correspondence.  This is
a straightforward matter of working out the structure of the appropriate
$(0,2)$ superfields and then repeating the procedure of \S3.

\section{$(0,2)$ Superfields}

\REF\hullo{C. Hull and E. Witten, ``Supersymmetric Sigma Models
And The Heterotic String,'' Phys. Lett. {\bf 160B} (1985) 398.}
\REF\ds{M. Dine and N. Seiberg, ``$(2,0)$ Superspace,'' Phys. Lett.
{\bf B180} (1986) 364.}
\REF\gateso{R. Brooks, J. Gates, and F. Muhammed, Nucl. Phys. {\bf B268}
(1986) 599, ``Extended $D=2$ Supergravity Theories And Their Lower
Superspace Realizations,'' Class. Quant. Grav. {\bf 5} (1988) 785.}
We will work in $(0,2)$ superspace, with bosonic coordinates
$y^\alpha,\,\,\alpha=1,2$, and fermionic coordinates $\theta^+,\b\theta^+$.
\foot{There have been previous discussions relevant to what follows
[\hullo,\ds,\gateso], but a few points made below are new and others
are reviewed for completeness.}
The supersymmetry generators are
$$\eqalign{
Q_+ & = {\partial\over\partial\theta^+}+i\b\theta^+\left({\p\over\p y^0}
+{\p\over\p y^1}\right) \cr
\b Q_+ & = -{\partial\over\partial\b\theta^+}-i\theta^+\left({\p\over\p y^0}
+{\p\over\p y^1}\right). \cr}          \eqn\buddo$$
These commute with
$$\eqalign{
D_+ & = {\partial\over\partial\theta^+}-i\b\theta^+\left({\p\over\p y^0}
+{\p\over\p y^1}\right) \cr
\b D_+ & = -{\partial\over\partial\b\theta^+}+i\theta^+\left({\p\over\p y^0}
+{\p\over\p y^1}\right), \cr}          \eqn\buddo$$
which are used in constructing Lagrangians.

I make no claim to describing here all possible $(0,2)$ models,
only those that can be conveniently described by certain types of
$(0,2)$ superfields.

\subsection{The Gauge Multiplet}

We want to introduce gauge fields in superspace.  The gauge covariant
derivatives will be called ${\cal D}_+$, $\b{\cal D}_+$, and ${\cal D}_\alpha
=D/D y^\alpha$.
We assume the constraints on the superspace gauge fields
$$\eqalign{ {\cal D}_+{}^2 & =\b{\cal D}_+{}^2 = 0 \cr
            {\cal D}_+\b{\cal D}_++\b{\cal D}_+{\cal D}_+ &
=2i(\cd_0+\cd_1).\cr
                  } \eqn\hjudu$$
(These equations, for instance, permit the existence of the $(0,2)$ chiral
superfields that we introduce later.)  The first two equations mean
that
$$\eqalign{ {\cal D}_+ & = e^{-\Psi}{ D}_+e^{\Psi} \cr
           \b{\cal D}_+& = e^{\b\Psi}\b{D}_+e^{-\b\Psi}, \cr}
\eqn\mmcc$$
with $\Psi$ a Lie algebra valued function.  By a gauge transformation
one can assume that $\Psi$ is real.  One can also gauge away
terms in $\Psi$ that are independent of $\theta^+ $ or of $\b\theta^+$.
In particular, one can go to an analog of Wess-Zumino gauge in which
$\Psi=\theta^+\b\theta{}^+(v_0+v_1)$, with $v_0+v_1$ a function of the
$y^\alpha$ only.  With this partial gauge fixing,
$$\eqalign{ \cd_0+\cd_1 & =\partial_0+\partial_1+i(v_0+v_1) \cr
            {\cal D}_+ & ={\p\over\p\theta^+}-i\b\theta^+(\cd_0+\cd_1) \cr
            \b{\cal D}_+ & =-{\p\over\p\b\theta^+}+i\theta^+(\cd_0+\cd_1).\cr}
                         \eqn\loporo$$
The rest of the superspace gauge field is
$$\cd_0-\cd_1=\partial_0-\partial_1 +iV, \eqn\oporo$$
with some function $V$ that can be expanded
$$V=v_0-v_1-2i\theta^+\b\lambda_--2i\b\theta^+\lambda_-
+2\theta^+\b\theta^+D.            \eqn\noporo$$
The transformation laws of $v_\alpha,\lambda_-,\b\lambda_-,$ and $D$
can be deduced from these formulas (using the supersymmetry
generators \buddo\ accompanied by gauge transformations to preserve
the form of \loporo) and are precisely the $(0,2)$ truncation of the
transformation laws \donkey\ of the $(2,2)$ gauge multiplet -- except
that some fields are missing.  The missing fields form a separate
$(0,2)$ multiplet that will be identified later.

The basic gauge invariant field strength is $\Upsilon=[\b\cd_+,\cd_0-\cd_1]$.
The natural superspace action for the $(0,2)$ gauge multiplet is
$$L_{\mit gauge}=
{1\over 8e^2}\int \d^2y \d\theta^+\d\b\theta^+\;\Tr \b\Upsilon\Upsilon
\eqn\dufo$$
This can be evaluated in components to give (in the $U(1)$ case, for
simplicity)
$${1\over e^2}
\int \d^2y\left({1\over 2}v_{01}{}^2+i\b\lambda_-(\partial_0+\partial_1)
\lambda_-+{1\over 2}D^2\right). \eqn\ofuo$$

\subsection{The Chiral Multiplet}

There are two types of matter multiplets to consider.  One is a bose
field $\Phi$, in some representation of the gauge group, obeying
$$\b \cd_+\Phi = 0. \eqn\hofo$$
We will call this the $(0,2)$ chiral multiplet.  The chiral multiplet
has a $\theta$ expansion
$$\Phi=\phi+\sqrt 2\theta^+\psi_+-i\theta^+\b\theta^+(D_0+D_1)\phi.\eqn\cofo$$
(Here $D_\alpha$ is now of course the ``ordinary'' gauge-covariant
derivative at $\theta^+=\b\theta{}^+=0$.)
If $\Phi$ is a field of charge $Q$, interacting with the abelian
gauge multiplet described in components above, the action is
$$\eqalign{
L_{\mit ch}=-{i\over 2}\int\d^2y\,\d^2\theta\,\,\,\,\b\Phi(\cd_0-\cd_1)\Phi
=\int \d^2y &
\left(-|D_\alpha \phi|^2+\b\psi_+i(D_0-D_1)\psi_+\right.
\cr &\left.-iQ\sqrt 2\b \phi
\lambda_-\psi_++iQ\sqrt 2\b\psi_+\b\lambda_-\phi+QD\b \phi \phi\right).\cr}
\eqn\foggyo$$

\subsection{The Fermi Multiplet}

The other type of matter multiplet is an anticommuting, negative
chirality spinor field $\Lambda_-$, in some representation of the
gauge group, obeying
$$\b{\cal D}_+\Lambda_- =\sqrt 2 E, \eqn\jokolo$$
where $E$ is some superfield obeying
$$\b{\cal D}_+E = 0 . \eqn\nokolo$$
We will call $\Lambda_-$ a fermi multiplet.
The $\theta$ expansion of the fermi multiplet is
$$\Lambda_-= \lambda_--\sqrt 2 \theta^+ G-i\theta^+\b\theta^+(D_0+D_1)\lambda_-
 -\sqrt 2\b\theta^+E. \eqn\booco$$
In turn $E$ has a theta expansion.  In the important case that
$E=E(\Phi_i)$ is a holomorphic function of some chiral superfields
$\Phi_i$ (with expansions as in
\cofo), one has
$$ E(\Phi_i)=E(\phi_i)+\sqrt 2\theta^+{\p E\over\p \phi_i}\psi_{+,i}
-i\theta^+\b\theta{}^+(D_0+D_1)E(\phi_i). \eqn\polyo$$

The natural action of the fermi multiplet is
$$L_F=-{1\over 2}\int\d^2 y\d^2\theta\;\b\Lambda_-\Lambda_-. \eqn\olyo$$
If $E$ is as in \polyo, then the component expansion is
$$L_F=\int \d^2y\left(i\b\lambda_-(D_0+D_1)\lambda_-+|G|^2
-|E(\phi_i)|^2 -\left(\b\lambda_-{\p E\over\p \phi_i}\psi_{+,i}
+{\p\b E\over\p\b \phi_i}\b\psi_{+,i}\lambda_-\right)\right) .\eqn\ucoor$$

\subsection{Other Terms In The Action}

Other terms in the action will be of the form
$$\int \d^2 y\d\theta^+\left.( \dots)\right|_{\b\theta^+=0}+{\mit h.c.},
\eqn\opory$$
where $\dots$ is some anticommuting superfield annihilated by $\b{\cal D}_+$.
These terms, which cannot be written as integrals over all of superspace,
are of particular importance.

For instance, the gauge field strength $\Upsilon$ obeys
$\b\cd_+\Upsilon=0$.  So in the abelian case
we can write
$$\eqalign{ L_{D,\theta}=&
{t\over 4}\int\d^2y\d\theta^+\left.\Upsilon\right|_{\b\theta^+=0}
+{\mit h.c.}\cr =&
{it\over 2}
\int\d^2y(D-iv_{01})-{i\b t\over 2}\int\d^2y(D+iv_{01}).
\cr}        \eqn\likko$$

For the other main example, let $\Lambda_{-,a}$ be some fermi superfields
with
$$\b{\cal D}_+\Lambda_{-,a}=\sqrt 2 E_a, \eqn\nikko$$
$E_a$ being some chiral superfields.  And let $J^a$ be some
chiral superfields with
$$ E_aJ^a=0. \eqn\orthorin$$
Then
$$\b{\cal D}_+(\Lambda_{-,a}J^a) =0. \eqn\northorin$$
So we can introduce another term in the action;
it is the $(0,2)$ analog of the superpotential:
$$L_{\mit J}=-{1\over \sqrt 2}
\int \d^2y\d\theta^+\left. \Lambda_{-,a}J^a\right|_{\b\theta^+=0}
-{\mit h.c.}\eqn\nurgo$$
If we suppose that $E_a$ and $J^a$ are holomorphic functions of
chiral superfields $\Phi_i$, we get
$$L_{\mit J}=-\int\d^2y\left(G_aJ^a(\phi_i)+\lambda_{-,a}\psi_{+,i}
{\p J^a
\over \p \phi_i}\right)-{\mit h.c.}\eqn\mucucc$$

Combining all this, we take the Lagrangian to be
$$L=L_{\mit gauge}+L_{\mit ch}+L_{\mit F}+L_{\mit D,\theta}+L_{\mit J}.
         \eqn\oturgo$$
After eliminating the auxiliary fields $D$ and $G_a$, the bosonic
potential turns out to be
$$U(\phi_i) = {e^2\over 2}\left(\sum_iQ_i|\phi_i|^2-r\right)^2 +\sum_a|E_a|^2
+ \sum_a|J^a|^2 .\eqn\hubonko$$

The $(0,2)$ analog of the C.Y./L.G. correspondence will be found
-- in a by now familiar fashion -- by studying the vacuum structure
as a function of $r$.

\subsection{Reduction Of $(2,2)$ Multiplets}

Now let us discuss the reduction of $(2,2)$ multiplets to
$(0,2)$ multiplets.

First, consider the $(2,2)$ gauge multiplet, described in Wess-Zumino
gauge by a scalar function $V(y,\theta^\pm,\b\theta^\pm)$
with the expansion of equation \urmo.  Part of
this multiplet makes up the $(0,2)$ gauge multiplet described
above.  The remaining fields are $\sigma,\lambda_+,$ and $\b\lambda_+$,
and make up a $(0,2)$ chiral multiplet.
In fact, this multiplet is simply $\Sigma'=\Sigma|_{\theta^-=\b\theta^-=0}$,
where $\Sigma=\sigma+\dots$ is the gauge invariant field strength of
the $(2,2)$ gauge multiplet, introduced in equation \huco.
\foot{To put the $\theta$ expansion of $\Sigma'$
 in the standard form \cofo\ for a $(0,2)$ chiral multiplet, one must
absorb a factor of $i$ in $\b\lambda_+$.}

Now consider a $(2,2)$ chiral multiplet $\Phi$.
$\Phi$ decomposes under $(0,2)$ supersymmetry into two multiplets.
First, there is a $(0,2)$ chiral multiplet
$$\Phi'=\left.\Phi\right|_{\theta^-=\b\theta^-=0}. \eqn\mocco$$
Second, let
$$\Lambda_-={1\over \sqrt 2}
\left. {\cal D}_-\Phi\right|_{\theta^-=\b\theta^-=0}.\eqn\occo$$
Then
$$\b{\cal D}_+\Lambda_-=
{1\over \sqrt 2}\left.\{\b{\cal D}_+,{\cal D}_-\}\Phi\right|_{\theta^-=
\b\theta^-=0}
.\eqn\moccoco$$
If, for instance, the gauge group is $U(1)$, and $\Phi$ is of charge $Q$,
this amounts to
$$\b{\cal D}_+\Lambda_- = 2iQ\Sigma'\Phi'.   \eqn\tococo$$
So $\Lambda_-$ is a fermi multiplet
with
$$ E=iQ\sqrt 2 \Sigma'\Phi'. \eqn\pococo$$

Finally, we want the $(0,2)$ reduction of the superpotential of
a $(2,2)$ model.  Consider a $(2,2)$ model with chiral superfields
$\Phi_i$ and a superpotential $W(\Phi_i)$.  Under $(0,2)$ supersymmetry,
the $\Phi_i$ split, as we have just seen, into $(0,2)$ chiral
superfields $\Phi'_i$, and fermi superfields $\Lambda_{-,i}$.
The latter obey $\b{\cal D}_+\Lambda_{-,i}=\sqrt 2 E_i$, where
from \pococo,
$$E_i=iQ_i\sqrt 2\Sigma'\Phi'_i. \eqn\kopo$$
While
in $(2,2)$ supersymmetry the scalar and Yukawa couplings of chiral
superfields are determined by a single function $W$, in $(0,2)$ supersymmetry
we must specify a collection of functions $J^i$, one for each $\Lambda_{-,i}$,
obeying $J^iE_i=0$, with $E_i$ given above. The $J^i$ that arise
in reduction of a $(2,2)$ model are simply
$$J^i={\partial W\over \partial \Phi_i'}. \eqn\bopo$$
This can be found by comparing equation \urgo\ for couplings derived
from a superpotential in $(2,2)$ models to
equation \mucucc\ for the couplings determined by the $J$'s.  With
$J^i$ as in \bopo\ and $E_i$ as in \kopo\ the equation $E_iJ^i=0$ is
a consequence of the gauge invariance of $W$.

\section{Some Models}

Now, let us apply this machinery to some sigma models with $(0,2)$
supersymmetry.
For simplicity, we will consider only the case in which the target
space is a hypersurface $X$ in $\CP^{n-1}$.

We know from \S3 which superfields we need to construct a super-renormalizable
$(2,2)$ model flowing at low energies to the sigma model of $X$.
For our present purposes, we must decompose those
superfields into $(0,2)$ multiplets, with a view to eventually
constructing $(0,2)$ deformations of the models of \S3 -- and their
Landau-Ginzburg description.

So we consider a $U(1)$ gauge theory in $(0,2)$ superspace, with
the following $(0,2)$ chiral multiplets: $n$ fields $S_i$ of
charge $1$,\foot{All supermultiplets here will be $(0,2)$ supermultiplets,
and I omit the primes from the notation.}
one field $P$ of charge $-n$, and one neutral field $\Sigma$ (from
the reduction of the $(2,2)$ gauge multiplet).  In addition, we want
$n$ fermi multiplets $\Lambda_{-,i}$, of charge $1$, with
$$E_i=i\sqrt 2 \Sigma S_i, \eqn\opurgo$$
and one more fermi multiplet $\Lambda_{-,0}$, of charge $-n$,
with
$$E_0=-in\sqrt 2 \Sigma P. \eqn\jopurgo$$

To complete the specification of the model, we need to pick additional
functions $J^i, \,\,i=1\dots n$, and $J^0$, with
$$E_iJ^i
+E_0 J^0=0.\eqn\hoko$$
In the $(2,2)$ case, we had the superpotential
$W=PG(S_i)$, with
$G$ a homogeneous $n^{th}$ order polynomial obeying the usual
transversality condition.  In $(0,2)$ language, this choice
according to \bopo\ leads to
$$\eqalign{ J^i & = P{\p G\over \p S_i} \cr
            J^0&  = G         .\cr}   \eqn\mino$$
To obtain $(0,2)$ models, we will change these choices while preserving
\hoko\ as well as gauge invariance.  We will also assume that
$J^0$ remains
independent of $P$ while $J^i$ remains linear in $P$.
(Other terms allowed by gauge invariance would be
of higher order in the chiral superfields
and apparently ``irrelevant'' in the renormalization
group sense.  The same is true of possible modifications of the $E$'s.)
So we may as well
preserve $J^0=G$ as the definition of $G$.  It is however possible
to change the $J^i$.  The general possibility is
$$\eqalign{ J^i & = P{\p G\over \p S_i}+PG^i \cr
            J^0& = G         ,\cr}   \eqn\mmino$$
where the $G^i$ are polynomials of degree $n-1$ in the $S_k$
chosen so that
$$S_iG^i = 0.  \eqn\olimmo$$

So we obtain a family of $(0,2)$ models parametrized, in addition to
the usual data, by the $G^i$.
Pulling together the relevant formulas, the potential energy of the
theory is
$$U(s_i,p)={e^2\over 2}\left(\sum_i|s_i|^2-n|p|^2-r\right)^2
+|G|^2+|p|^2\sum_i\left|{\p G\over \p s_i}+G^i\right|^2
+2|\sigma|^2\left(\sum_i |s_i|^2+n^2|p|^2\right). \eqn\cocnon$$
Here the bosonic components of $S_i,P$, and $\Sigma$ have been
called $s_i$, $p$, and $\sigma$.
We can therefore analyze the $r$ dependence just as in the $(2,2)$
case.  For $r>>0$ we get a phase in which the low energy theory is
a $(0,2)$ sigma model with target space $X$.  For $r<<0$, we get a
$(0,2)$ Landau-Ginzburg phase.
We have generalized the C-Y/L-G correspondence to $(0,2)$ models.
Let us look at the two phases more closely.

\subsection{The Landau-Ginzburg Phase}

For $r<<0$, $p$ gets an expectation value, just as in the $(2,2)$ case,
breaking the gauge group to $\Z_n$.  The model is therefore
a $\Z_n$ orbifold.

The massless multiplets are the chiral multiplets $S_i$ and $\Lambda_{-,i}$;
other multiplets are massive.  The effective values of $E_i$ and $J^i$
in the low energy theory are obtained by setting
$\sigma$ and $p$ to their vacuum expectation values, namely 0 and
$\sqrt{-r/n}$.  So one has in the effective $(0,2)$ Landau-Ginzburg model
$$\eqalign{
  E_i & = 0 \cr
  J^i & = <p>\left({\p G\over \p s_i}+ G^i\right). \cr}\eqn\mikolo$$
Of course these obey $E_iJ^i=0$.

\subsection{The Calabi-Yau Phase}

For $r>>0$ the massless bose modes are the modes tangent to $X$.
The right-moving massless fermions are just
the $(0,2)$ partners of the massless bosons.

But what about left-moving massless fermions?  Here the situation
is more interesting.  The fermi component of $\Lambda_{-,0}$ is massive,
but certain modes of $\lambda_{-,i}$ (the fermi components of the superfields
$\Lambda_{-,i}$) are massless.  By inspection of the above formulas for
the Lagrangian,
the massless modes can be seen to obey
$$\eqalign{\sum_i\b s_i\lambda_{-,i} & = 0 \cr}\eqn\nunko$$
and also
$$\eqalign{
       \sum_i\left({\p G\over \p s_i}+G^i \right)\lambda_{-,i} & = 0 . \cr}
   \eqn\hunko$$
Let us put these formulas in the general format of $(0,2)$ sigma models.
In general, massless left-moving fermions are sections of some holomorphic
vector bundle $F$ over the target space $X$. This means that it must
be possible to describe the massless modes by equations that vary
holomorphically in the $s_i$.  Now, \hunko\ is holomorphic in $s_i$,
but \nunko\ is not.  To achieve more understanding, we should replace
\nunko\ by an equation that has the same consequences but varies
holomorphically
in $s$.  This is easily done.  Instead of the constraint \nunko, introduce
a gauge invariance, saying that two modes of $\lambda_i$ are considered
equivalent if they are related by a transformation
$$\lambda_i\to \lambda_i + s_i\lambda, \eqn\gunko$$
for some $\lambda$.  Then \nunko\ can be regarded as a gauge fixing
condition; each orbit of the invariance \gunko\ has a unique representative
obeying \nunko.

To put \gunko\ and \hunko\ in their theoretical context,
introduce the usual line bundles
${\cal O}(k)$ over $\CP^{n-1}$, restricted to $X$, and  consider
a sequence of maps of vector bundles over $X$,
$$0\to {\cal O}\underarrow{\alpha} \oplus_{i=1}^n{\cal O}(1)
  \underarrow{\beta} {\cal O}(n)\to 0, \eqn\funko$$
with $\alpha$ being the map $\lambda\to s_i\lambda$, and
$\beta$ the map
$$\lambda_i \to \lambda_i\left({\p G\over \p s_i}+G^i\right).\eqn\micoo$$
\funko\ is a complex of vector bundles (that is $\beta\alpha=0$), but it
is not an exact sequence.  On the contrary, ${\rm ker}\,\beta/{\rm im}\,\alpha$
is a rank $n-2$ holomorphic vector bundle $F$ over $X$.

\REF\sgw{M. B. Green, J. H. Schwarz, and E. Witten, {\it Superstring Theory},
vol. 2 (Cambridge University Press, 1987).}
For $G^i=0$,
$F$ reduces to the tangent bundle $TX$ of $X$.  In general, $F$ is a
deformation of $TX$, and conversely, all deformations of $TX$ as a
holomorphic vector bundle over $X$ can be described as in \funko.
This  family of deformations of $TX$
has been discussed previously [\sgw, volume II, pp. 461-3].
The choice of $F$ (together with the rest of the data)
determines a $(0,2)$ model with target $X$,
so  the $G^i$ parametrize a family of such models;
we have found the Landau-Ginzburg description of that family.

One may wonder what concrete results
follow from this $(0,2)$ analog of the usual C-Y/L-G correspondence.
One answer, along the lines of \S3.3, is that $\Tr(-1)^F$ and
the elliptic genus are invariant under the transition from Calabi-Yau
to Landau-Ginzburg models, while for the half-twisted $(0,2)$ model
(which determines low energy Yukawa couplings) Landau-Ginzburg is
an analytic continuation of Calabi-Yau.
(The $A$ and $B$ topological field theories do not have $(0,2)$ analogs.)
Another answer, along the lines
of \S5.5, is that topology-changing processes involving $(0,2)$ models
can be analyzed just as for $(2,2)$ models.

To complete the story, perhaps I should point out that in $(0,2)$ sigma
models with left-moving massless fermions taking values in some vector
bundle $F$ over the target space $X$, the objects $E_a$ and $J^a$
are geometrically to be understood as holomorphic sections of $F$ and of $F^*$
(the dual of $F$), respectively.  If non-zero, these sections give
masses to some of the modes.  In the particular example we have
been studying with $X$ a hypersurface in $\CP^{n-1}$ and $F$ as above,
the effective $E_a$ and $J^a$ are 0.  They could hardly
be otherwise, as in these examples $F$ and $F^*$ have no non-zero
global holomorphic
sections.  But in the general study of $(0,2)$ sigma models, when
$F$ is such that $F$ or $F^*$ have global sections, the consideration
of $E_a$ and $J^a$ is likely to be of great importance.

\subsection{Conformal Invariance?}

In \S3, we constructed the C-Y/L-G correspondence for $(2,2)$ models
without having to know whether conformal invariance is valid on either
side.  In the present section, we have done the same for $(0,2)$ models.

\REF\greene{J. Distler and B. Greene, ``Aspects Of $(2,0)$ String
Compactifications,'' Nucl. Phys. {\bf B304} (1988) 1.}
For the $(0,2)$ models, the situation is believed to be
quite different from that for $(2,2)$ models.
The $(2,2)$ models studied in \S3 are all believed to have conformally
invariant fixed points to which they flow in the infrared.
Order by order in perturbation theory, the same is true for the
$(0,2)$ models analyzed above.
But in many cases, these
approximate $(0,2)$ conformal fixed points are destabilized by nonperturbative
corrections [\dine,\greene].  For instance, for the particular
case we have been looking at closely, it is believed that this
occurs for generic choices of the polynomials $G,G^i$.
Little is understand, from the world-sheet point of view, about the
nature of the breakdown of conformal invariance.

We have, for illustrative purposes, studied in detail only $(0,2)$
models that can be constructed as perturbations of $(2,2)$ models.
It would be extremely interesting to study other $(0,2)$ models
using the techniques in this paper.
\ack{I am grateful to P. Aspinwall, B. Greene, and D. Morrison for
illuminating discussions
and for describing their work at a preliminary stage.}
\refout
\end